\documentclass[11pt]{article}
\usepackage{jheppub}
\usepackage[english]{babel}
\usepackage[numbers,sort&compress]{natbib}
\usepackage{amsmath,amssymb,amsbsy,amstext,amsthm,booktabs,simplewick,exscale,relsize,slashed,graphicx,amsfonts,upgreek}
\usepackage{multirow,dcolumn,bm,enumerate,url,hyperref}

\def\tablename{Table}
\def\figurename{Figure}

\newcommand\one{\leavevmode\hbox{\small1\normalsize\kern-.33em1}}

\newcommand{\qqquad}{\qquad \qquad}

\newcommand{\met}{\slashchar{p}_T}

\newcommand{\mev}{\text{MeV}}
\newcommand{\gev}{\text{GeV}}
\newcommand{\tev}{\text{TeV}}
\newcommand{\fb}{\text{fb}}

\newcommand{\pb}{\text{pb}}

\newcommand{\iab}{\text{ab}^{-1}}

\def\slashchar#1{\setbox0=\hbox{$#1$}           % set a box for #1
   \dimen0=\wd0                                 % and get its size
   \setbox1=\hbox{/} \dimen1=\wd1               % get size of /
   \ifdim\dimen0>\dimen1                        % #1 is bigger
      \rlap{\hbox to \dimen0{\hfil/\hfil}}      % so center / in box
      #1                                        % and print #1
   \else                                        % / is bigger
      \rlap{\hbox to \dimen1{\hfil$#1$\hfil}}   % so center #1
      /                                         % and print /
   \fi}

\newcommand{\eg}{\textsl{e.g.}\;}
\newcommand{\ie}{\textsl{i.e.}\;}

\newcommand{\be}{\begin{eqnarray*}}
\newcommand{\ee}{\end{eqnarray*}}

\newcommand{\bee}{\begin{eqnarray}}
\newcommand{\eee}{\end{eqnarray}}
\newcommand{\beeq}{\begin{equation}}
\newcommand{\eeeq}{\end{equation}}

\begin{document}
\begin{flushright}
FERMILAB-PUB-14-517-T\\
IPMU14-0346\\
\end{flushright}
\title{The Relic Neutralino Surface at a 100 TeV collider}

\author[a]{Joseph Bramante,}
\author[b]{Patrick J. Fox,}
\author[a]{Adam Martin,}
\author[a]{Bryan Ostdiek,}
\author[c]{Tilman Plehn,}
\author[c]{Torben Schell,} 
\author[d]{and Michihisa Takeuchi}

\affiliation[a]{Department of Physics, University of Notre Dame, IN, USA}
\affiliation[b]{Theoretical Physics Department, Fermilab, Batavia, IL USA}
\affiliation[c]{Institut f\"ur Theoretische Physik, Universit\"at Heidelberg, Germany}
\affiliation[d]{Kavli IPMU (WPI), The University of Tokyo, Kashiwa, Japan}

\abstract{We map the parameter space for MSSM neutralino dark matter
  which freezes out to the observed relic abundance, in the limit that
  all superpartners except the neutralinos and charginos are 
  decoupled. In this space of relic neutralinos, we show the
  dominant dark matter annihilation modes, the mass splittings among the electroweakinos, direct detection
  rates, and collider cross-sections.\footnote{Animated 3-dimensional versions of the relic neutralino surface are available from \url{http://www3.nd.edu/~bostdiek/research_welltmp.html}.} 
 The mass difference between the
  dark matter and the next-to-lightest neutral and charged
  states is typically much less than electroweak gauge boson masses.
  With these small mass differences, the relic neutralino surface is accessible to a future
  100 TeV hadron collider, which can discover inter-neutralino mass
  splittings down to 1~GeV and thermal relic dark matter neutralino
  masses up to 1.5 TeV with a few inverse attobarns of luminosity. This
  coverage is a direct consequence of the increased collider energy: in the Standard Model
  events with missing transverse momentum in the TeV range have mostly hard electroweak radiation,
  distinct from the soft radiation shed in compressed electroweakino decays. We
  exploit this kinematic feature in final states including photons
  and leptons, tailored to the 100 TeV collider environment.}

\maketitle

\setcounter{footnote}{0}

%%%%%%%%%%%%%%%%%%%%Section Introduction%%%%%%%%%%%%%%%%%%%%%%%%%
\section{Introduction}
\label{sec:intro}
%%%%%%%%%%%%%%%%%%%%Section Introduction%%%%%%%%%%%%%%%%%%%%%%%%%

Understanding the properties of dark matter is a next major step in
experimental and theoretical particle
physics~\cite{Morrissey:2009tf}. In this paper we will establish that
a $100~\tev$ collider provides excellent prospects for detecting weakly interacting dark
matter. In addition, we show that a $100~\tev$ collider can expose nearby states that interact with the
dark matter.

Many scenarios of physics beyond the Standard Model predict that
dark matter is the lightest particle charged under a stabilizing
symmetry, and that it freezes out during the radiation-dominated
expansion of the universe.  In this paper we focus on the neutralino 
and chargino sector of the minimal supersymmetric
Standard Model (MSSM), where the lightest neutralino freezes out to
the observed relic abundance.  We refer to the four neutralinos and
the two charginos of the MSSM collectively as electroweakinos, and the
lightest neutralino as the lightest supersymmetric particle
(LSP). Over the supersymmetric parameter space this Majorana fermion
can be a mixture of the neutral components of a triplet under
$SU(2)_L$ (a wino), a singlet (a bino), or two $SU(2)_L$ doublets
(higgsinos). If the supersymmetric mass parameters of the
electroweakinos are well separated,
inter-multiplet mixing can be neglected and
the lightest neutralino can be studied as a pure state: pure singlet,
triplet, or doublet, depending on which supersymmetric mass parameter
is lowest. However, once we apply LEP, LHC, and astrophysical
constraints, the only pure state possibility that 
can fit
the required relic abundance is pure higgsino~\cite{lepii,Cohen:2013ama,Fan:2013faa,Aad:2013yna}.  There is
much more viable parameter space if we give up the pure state
hypothesis and allow the lightest neutralino to be an admixture of
bino, wino and higgsino, so-called
well-tempering~\cite{ArkaniHamed:2006mb}. In both the pure higgsino or
well-tempered scenarios, there are additional neutralino and chargino
states that have similar mass to the LSP. These states play an
important role in establishing the observed dark matter density, and
they will be crucial to the collider studies proposed here.

While the electroweakino sector of the MSSM is just one example of a
dark matter framework, the existence of other nearby (in mass) states
that communicate with the dark matter at the renormalizable level can
be argued from fairly general grounds. To understand why, let us
assume dark matter is a thermal relic and has some non-gravitational
interactions with the Standard Model. The partial wave amplitude unitarity bound
in Ref.~\cite{Griest:1989wd} implies that any dark matter populating
the universe through the classic freeze-out mechanism is lighter than
$340~\tev$, so we can break up the models of dark matter into two groups:
$m_{\chi} < 340~\tev$ and $m_{\chi} > 340~\tev$, where $m_{\chi}$ is
the mass of the dark matter agent.

For dark matter lighter than $340~\tev$, the weakly interacting massive
particle (WIMP) and WIMPless
miracles~\cite{Chiu:1966kg,Steigman:1979kw,Scherrer:1985zt,Feng:2008ya}
emphasize that if the correct dark relic abundance is attained through
freeze-out via a single mediator, then
\begin{alignat}{5}
\Omega_D h^2 = \frac{1}{\left\langle \sigma_a v \right\rangle} = \frac{m_D^2}{g_D^4} \simeq \frac{1}{\text{picobarn}} \; ,
\end{alignat}
where ($m_D,g_D$) are the mass and coupling associated with dark
matter's thermal annihilation cross-section, $\left\langle \sigma_a v
\right\rangle$. However, direct detection experiments constrain both
spin-independent and spin-dependent dark matter cross-sections with
Standard Model particles to be well below a picobarn for dark matter
masses less than
$100~\tev$~\cite{Xiao:2014xyn,Agnese:2014aze,Akerib:2013tjd,Aprile:2012nq,Aprile:2013doa}. Altogether,
this implies that either a new mass state mediates dark matter
freeze-out, or a nearby mass state allows for non-standard dark matter
annihilation\footnote{There are, of course, exceptions to the standard freeze-out scenario, e.g. freeze-in~\cite{Hall:2009bx} and asymmetric dark matter~\cite{Zurek:2013wia, Petraki:2013wwa}, and exceptions to constraints on dark matter-Standard Model cross-sections, e.g. leptophilic dark matter~\cite{Fox:2008kb}.}.

For dark matter heavier than $340~\tev$ a viable relic abundance can
still be achieved if the dark matter mass lies near a cross-section
pole (\eg dark matter exactly half the mass of a new $s$-channel mediator) or if it lies
below nearly mass-degenerate states that induce additional,
co-annihilation interactions~\cite{Griest:1990kh}. In both of these
cases, the lightest mass eigenstate would normally over-populate the
primordial relic abundance, but instead annihilates to the observed
abundance as a result of other states in the thermal bath.

The necessity of such an extended dark matter `sector', rather than a
single dark matter particle, makes resolving dark matter mass
splitting vital in parsing the physical implications of a detection of
dark matter, wherever it occurs. In particular for collider searches
this emphasizes the need to resolve nearby states, $O(10\%)$ heavier than
the primary dark relic.

The existence of an extended dark sector is certainly true in minimal supersymmetric models where the lightest neutralino is the LSP and a cosmologically viable relic abundance is achieved. The neutralino LSP is always accompanied by some similar-mass states, usually within $\mathcal{O}(10\%)$ of the LSP mass, allowing for co-annihilation during freeze-out. This happens both in constrained models~\cite{Henrot-Versille:2013yma,Bechtle:2012zk,Cohen:2013kna,Fowlie:2012im,Strege:2012bt,Bhattacherjee:2013vga,Buchmueller:2013rsa,Roszkowski:2014wqa} and the full
MSSM~\cite{Henrot-Versille:2013yma,Boehm:2013qva,Fowlie:2013oua,Cahill-Rowley:2013vfa,Cahill-Rowley:2014boa,Chakraborti:2014gea,Roszkowski:2014iqa}, though exactly which states are nearby can vary greatly. Sleptons,
squark, or charginos are all possibilities~\cite{Roszkowski:1991ng,Mizuta:1992qp,Edsjo:1997bg,Ellis:1998kh,Ellis:1999mm,Feng:2000gh,Ellis:2001nx,BirkedalHansen:2002am,Nihei:2002sc}. Squarks are far easier to produce at hadron colliders than electroweak particles, and are more strongly constrained by LHC limits. Sleptons are also fairly constrained due to their clean decay signal. The most difficult scenario to detect, and thus the focus of our work, is where the dark sector is purely composed of the electroweakinos (mass $0.2-3~\tev$) and all other supersymmetric particles have been decoupled ($\gtrsim 10~\tev$). 

In addition to being difficult to constrain at the LHC, the interactions between electroweakino sector dark matter and nuclei are often extremely suppressed.
Limits on the spin-dependent scattering of WIMPs
\cite{Behnke:2010xt,Felizardo:2011uw,Archambault:2012pm,Aprile:2013doa}
and indirect searches for MSSM electroweakino annihilation in the sun
\cite{Tanaka:2011uf,Aartsen:2012kia}, at best constrain dark matter
spin-dependent nucleon scattering to be less than
$10^{-40}~\rm{cm^2}$, but as we will see, the spin-dependent
cross-section of relic electroweakino dark matter can be $\lesssim
10^{-45}~\rm{cm^2}$. In addition, while limits on spin-independent
scattering
\cite{Xiao:2014xyn,Agnese:2014aze,Akerib:2013tjd,Aprile:2012nq} bound
the dark-matter nucleon cross-section to be smaller than
$10^{-44}~\rm{cm^2}$ for TeV-mass dark matter, Section
\ref{sec:paramspace} will show that relic bino-wino dark matter has
blind spots where the (tree-level) spin-independent cross section is
smaller than $10^{-50}~\rm{cm^2}$. Indeed, it was recently shown at next-to-leading
order that if a doublet, triplet, singlet-doublet, or doublet-triplet
of $SU(2)_L$ is heavy compared to a typical direct detection momentum
transfer, the absolute spin-independent scattering cross-section of
such a particle will be around $
10^{-48}~\text{cm}^2$~\cite{Hisano:2011cs,Hisano:2012wm,Hill:2013hoa,Hill:2014yka,Hill:2014yxa}.

For TeV-mass dark matter, this cross-section is much smaller than the solar
neutrino scattering cross-section, which provides a major irreducible
background for current xenon and semiconductor-based direct detection
methods~\cite{Cushman:2013zza}. Future directional dark matter
detection experiments may overcome the solar neutrino background by
subtracting off the solar neutrino diurnal variation
Refs.~\cite{Mayet:2011nk,Alves:2012ay,Mayet:2013mpa}. But even
assuming these directional methods advance to their full potential and
we find dark matter in low momentum transfer regimes, this will only
be a first step in illuminating the properties of dark
matter. Multiple detector materials across a number of experiments~\cite{Fox:2010bz,Fox:2010bu,Frandsen:2011gi,HerreroGarcia:2012fu,Gondolo:2012rs,Fox:2014kua} 
would be necessary to fully characterize even the mass of a
WIMP~\cite{McDermott:2011hx,Peter:2013aha,Gluscevic:2014vga}. This suggests that production of dark matter
at a hadron collider will be essential in clarifying any findings of
direct detection experiments. While a low momentum transfer detection of dark matter could occur before collider
detection, exposing the structure of the dark sector will require collider input.

Altogether, future direct and indirect detection are likely to leave
one well-motivated scenario for physics beyond the Standard Model largely
untested, namely a universe containing only a few weakly interacting
fermions in the $0.2-4~\tev$ mass range. In fact, this scenario realized in
Split Supersymmetry has been proposed as a parsimonious route to GUT scale
unification, albeit with a necessarily fine-tuned solution to the
hierarchy problem~\cite{Wells:2003tf,ArkaniHamed:2004fb,Giudice:2004tc,Wells:2004di,Kilian:2004uj,Arvanitaki:2012ps,Hall:2012zp,Unwin:2012fj,Kahn:2013pfa,Lu:2013cta,Fox:2014moa,Nagata:2014wma,Nomura:2014asa}. Regardless of UV motivations, if dark matter is a relic neutralino, there are
significant blind spots in electroweakino mass parameter space where
spin-independent and spin-dependent scattering (through LSP coupling
to the Higgs and $Z$) vanish, as examined in
Refs.~\cite{Cheung:2012qy,Han:2013gba,Cheung:2013dua,Huang:2014xua,Han:2014nba}. 
We will see that most of the presently unplumbed pockets of this MSSM
parameter space can be unmasked by $100~\tev$ collider
searches. 

Recently, there has been significant work motivating a
$100~\tev$ collider~\cite{Yu:2013wta,Anderson:2013ida,Cohen:2013xda,Curtin:2014jma,Fowlie:2014awa,Rizzo:2014xma,Larkoski:2014bia,Hook:2014rka,Barr:2014sga}
from a dark matter
perspective~\cite{Low:2014cba,Cohen:2014hxa,Cirelli:2014dsa,Acharya:2014pua,Gori:2014oua,Curtin:2014cca,diCortona:2014yua}. Key
analysis strategies are mono-jets searches, soft leptons, and
disappearing tracks. For pure wino and pure higgsino dark matter in
the MSSM, disappearing tracks are a promising search strategy as long
as the mass differences between the lightest electroweakino states are
perturbatively stable at the $200~\mev$ level.  Mixed relic
neutralinos with additional heavier states can be targeted with soft
(tri-)leptons~\cite{Low:2014cba}. Same-sign or opposite-sign dilepton
searches and tri-lepton signatures have the potential to cover some
neutralino parameter space, based on inclusive effective mass and
transverse momentum cuts~\cite{Gori:2014oua}. The additional
production of neutralinos and charginos in weak boson
fusion~\cite{Cho:2006sx} will be rate limited and only minimally
contribute to the discovery potential~\cite{Cirelli:2014dsa}. In this
paper we add a lepton--photon decay signature, which targets
electroweakino mass splittings above $1~\gev$ and covers similar
parameter space as the disappearing tracks, but without their
sensitivity to a tuned mass splitting. 

Moreover, the existing studies of electroweakino production at a
$100~\tev$ collider focus on pure electroweakino states and do not 
take cosmological aspects into account. In this work, we link
electroweakino discovery prospects directly to cosmological relic
abundance requirements. 
As a starting point, we survey MSSM neutralino dark matter mass
parameters to $4~\tev$ and chart every combination thereof which
freezes out to the correct relic dark matter abundance. Most of this parameter space surface of relic neutralinos is
accessible at a $100~\tev$ hadron collider, making use of the
production of slightly heavier electroweakinos and subsequent decays
through emission of an off-shell $Z \to \ell^+ \ell^-$ or a
photon. As we will see, the combination of $\sim \tev$ of missing energy with low momenta ($\sim5-50~\gev$) photons and/or leptons provides a sharp signature at very high energy hadron colliders.  

%%%%%%%%%%%%%%%%%%%%Section Parameter Space%%%%%%%%%%%%%%%%%%%%%%%%%
\section{The relic neutralino surface}
\label{sec:paramspace}
%%%%%%%%%%%%%%%%%%%%Section Parameter Space%%%%%%%%%%%%%%%%%%%%%%%%%

Neutralinos in the MSSM are mixed mass eigenstates of the partners of
the hypercharge gauge boson, the $SU(2)_L$ gauge boson, and Higgs
bosons. This means that, depending on their field content, the lightest
mass eigenstate interpolates between different WIMP scenarios. We
begin with an exploration of parameter space defined by the spectrum
of Standard Model fields plus electroweakino fields, with the added
requirement that the lightest neutral electroweakino forms the dark
matter relic abundance by freezing out of the primordial thermal
bath. We decouple all non-electroweakino MSSM fields, \ie the squarks, sleptons, and additional Higgs bosons. 
With these assumptions, we have 4 free parameters, the
mass of the bino ($M_1$), the mass of the wino ($M_2$), the mass of
the higgsinos ($\mu$), and the ratio of the vacuum expectation values
of the two Higgs fields ($\tan~\beta$). After fixing $\tan~\beta$ we
will study where in this 3-dimensional parameter space the lightest
neutralino freezes out to the observed relic density, its 
spin-dependent and spin-independent nucleon cross-sections, 
what the mass difference between the lightest and second lightest
electroweakino is, and what the relevant $100~\tev$ hadron collider cross-sections
are. In particular, the absolute mass splitting between the LSP dark
matter agent and the neutralino or chargino NLSP will be relevant for the
$100~\tev$ collider search strategy. 

Assuming thermal freeze-out of the lightest neutralino, the relic
abundance depends on the neutralino's mass, annihilation
cross-section, and the mass of any nearly mass-degenerate co-annihilating
particles. If the LSP is a pure gauge eigenstate with a non-trivial
$SU(2)_L$ representation, for instance the higgsino or the wino, these
will need to have large masses to compensate for efficient
annihilations depleting their number density during the
radiation-dominated era.  In the MSSM the masses of the lightest pure
WIMP states have to exceed $1~\tev$ for the higgsino and $3~\tev$ for the
wino.  In contrast, the gauge singlet bino inefficiently annihilates
and will over-close the universe if its mass is above $\sim
5~\gev$. If the lightest mass eigenstate mixes binos, higgsinos, and
winos, either through explicit fractions of the lightest mass
eigenstate or if the lightest eigenstate co-annihilates with heavier
electroweakinos, the resultant relic neutralinos can yield the
observed relic abundance for masses very different than in the case of pure
neutralino gauge eigenstates.

%---------------------------------------
\begin{figure}[t]
\begin{center}
\includegraphics[width=0.49\textwidth]{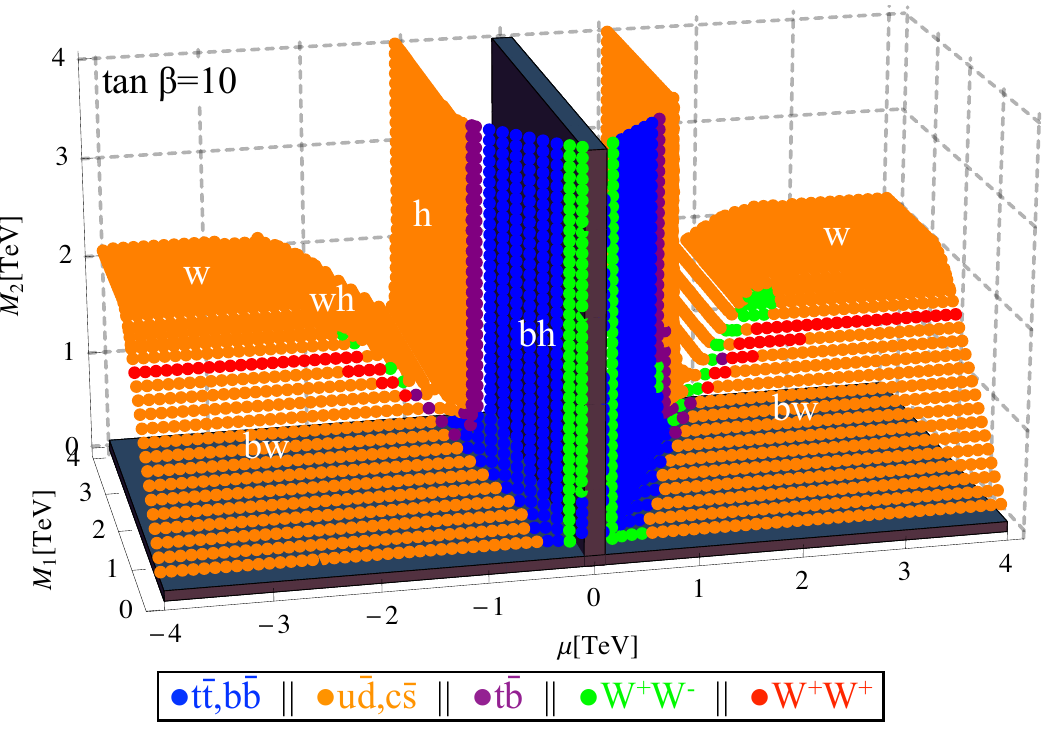}
\includegraphics[width=0.49\textwidth]{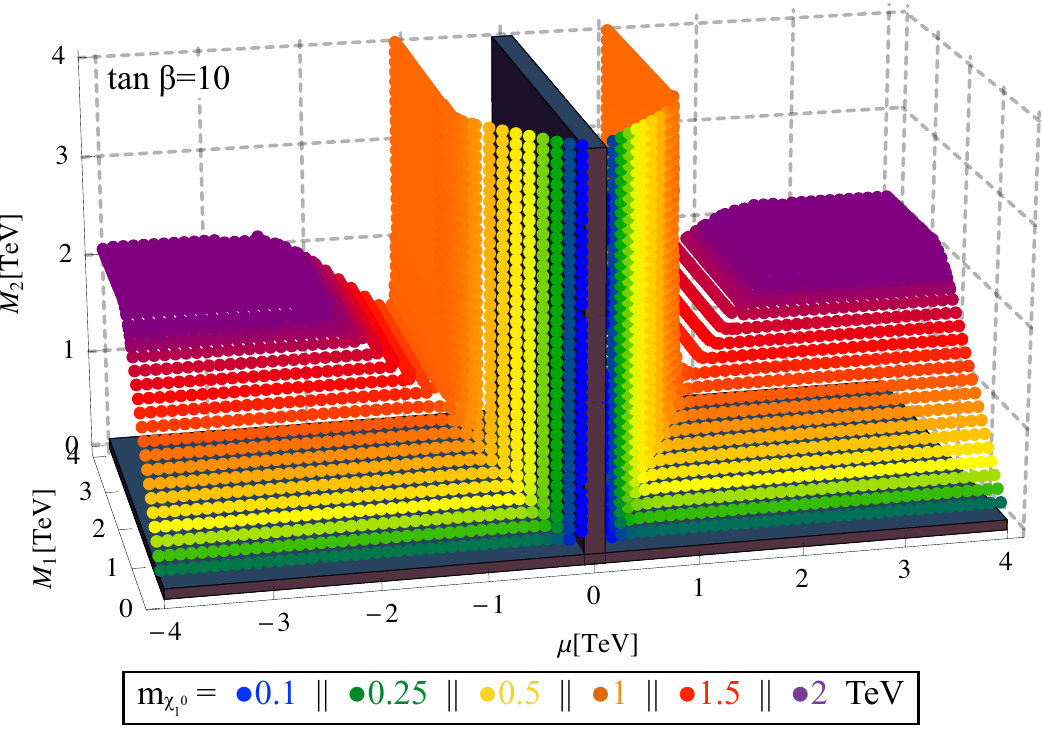}
\caption{Left panel: relic neutralino surface with the largest fraction of
  primordial annihilation products indicated by color. All points
  shown predict a dark matter relic density of $\Omega h^2 \simeq
  0.12$. Regions ruled out by LEP constraints are occluded with a dark
  box.  Planar surfaces of relic higgsinos, winos, bino--winos,
  bino-higgsinos, and wino-higgsinos are indicated with white
  letters. Right panel: mass of the lightest neutralino, the LSP, in TeV.}
\label{fig:surface_ann}
\end{center}
\end{figure}  
%---------------------------------------

\begin{figure}[t]
\includegraphics[width=0.49\textwidth]{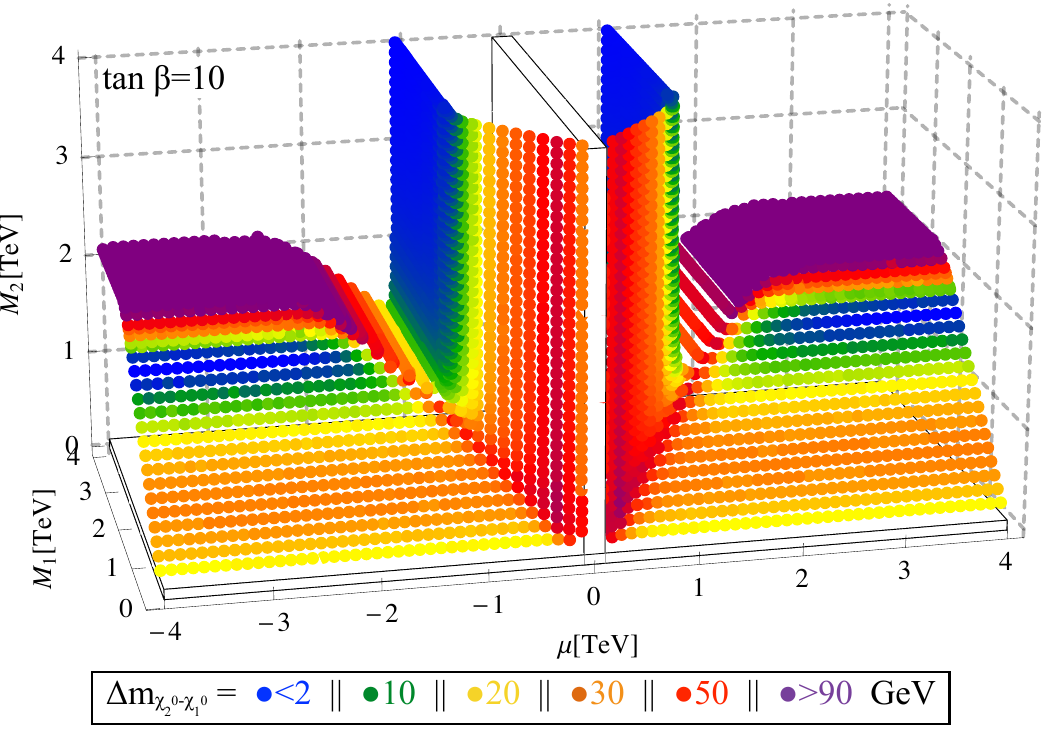} 
\includegraphics[width=0.49\textwidth]{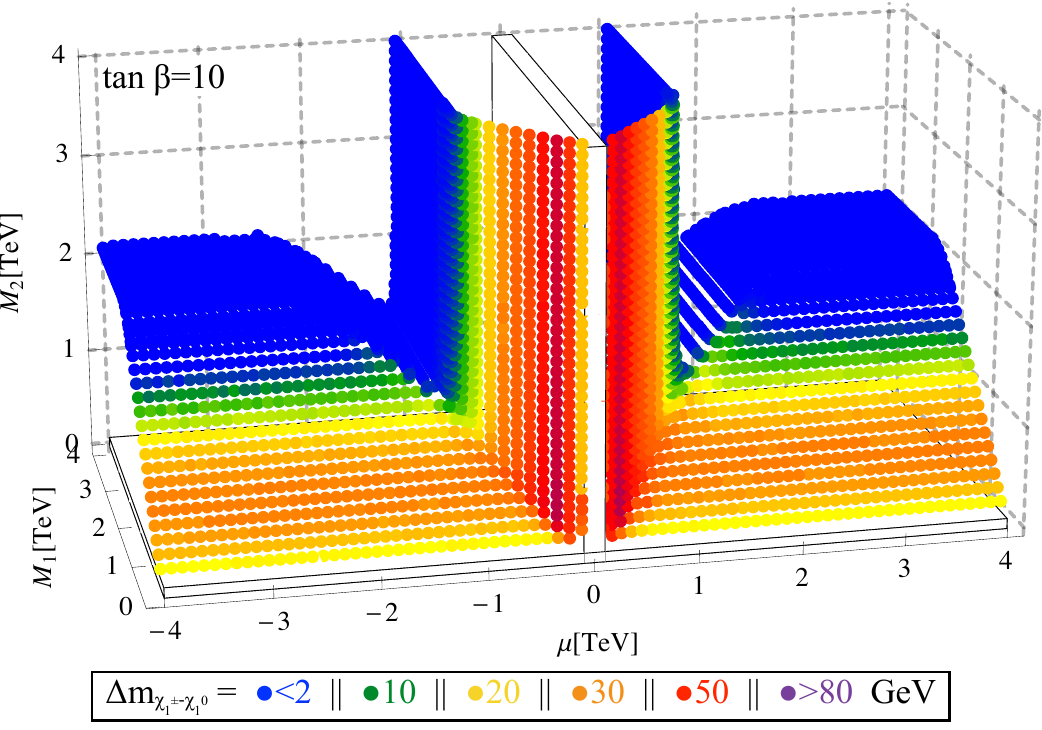}
\caption{Relic neutralino surface with mass splitting between
  $\chi_1^0$ and $\chi_2^0$ (left) and between $\chi_1^0$ and
  $\chi_1^\pm$ (right) as indicated. Regions ruled out by LEP are
  occluded with a white box.}
\label{fig:surface_split}
\end{figure}

\begin{figure}[t]
\includegraphics[width=0.49\textwidth]{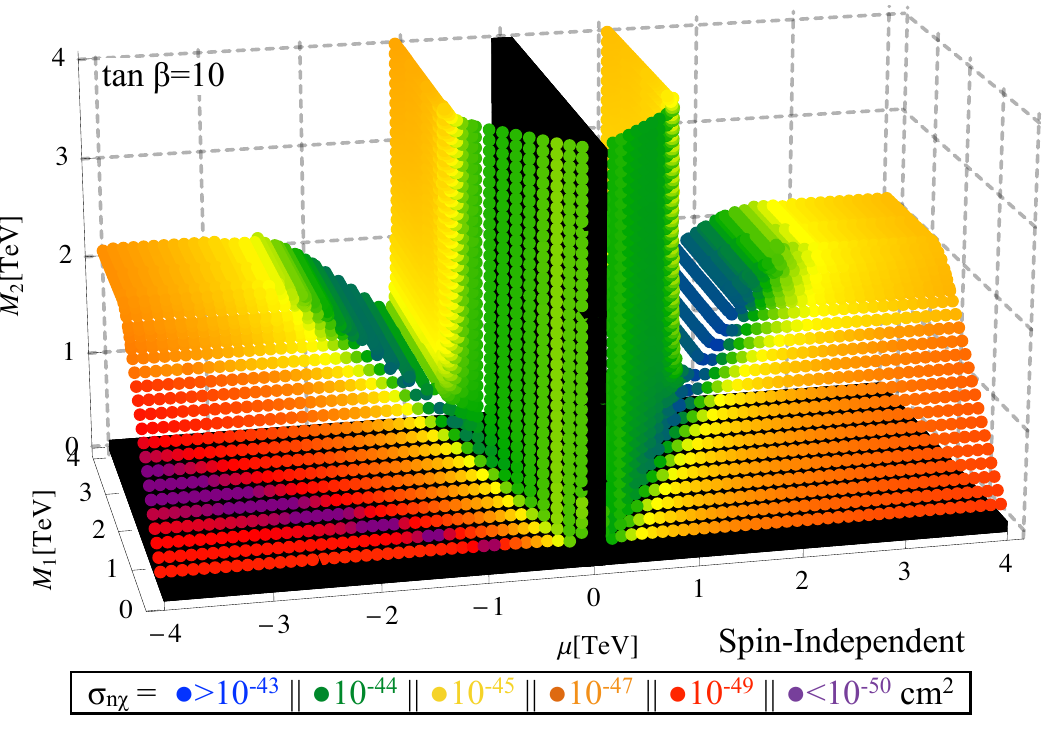} 
\includegraphics[width=0.49\textwidth]{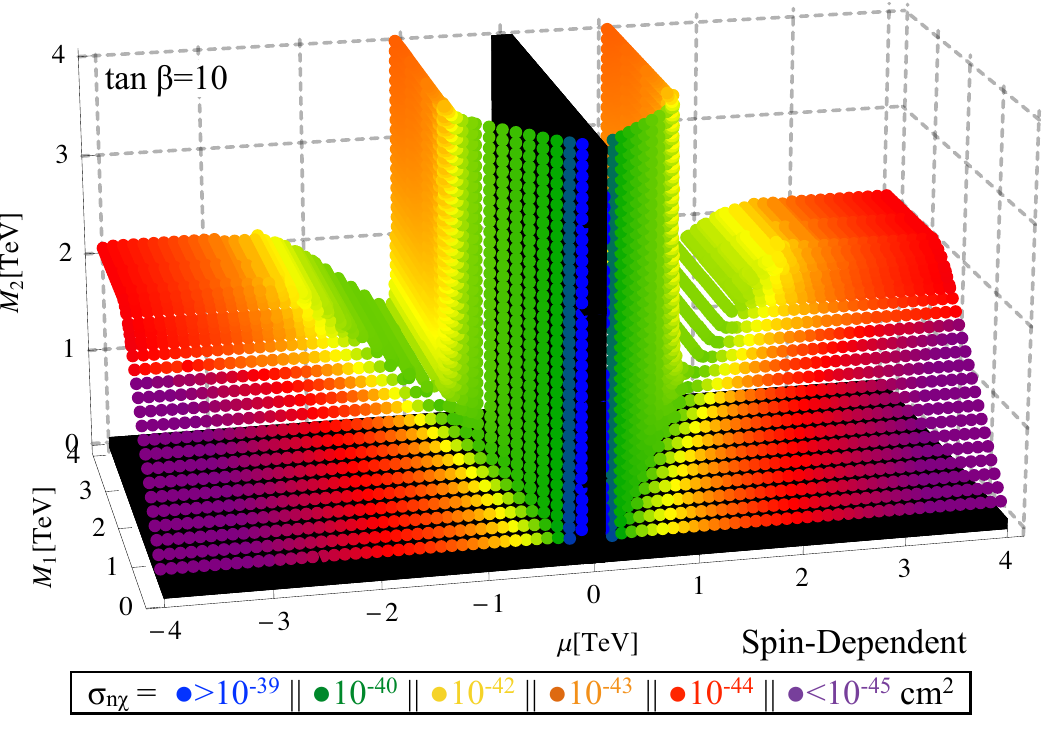}
\caption{Relic neutralino surface with spin-independent (left) and
  spin-dependent (right) scattering cross-sections of the LSP on
  nucleon. At each point in each plot, the larger of the $\chi$-proton and
  $\chi$-neutron scattering cross-sections is given.}
\label{fig:surface_dd}
\end{figure}

\begin{figure}[t]
\includegraphics[width=0.49\textwidth]{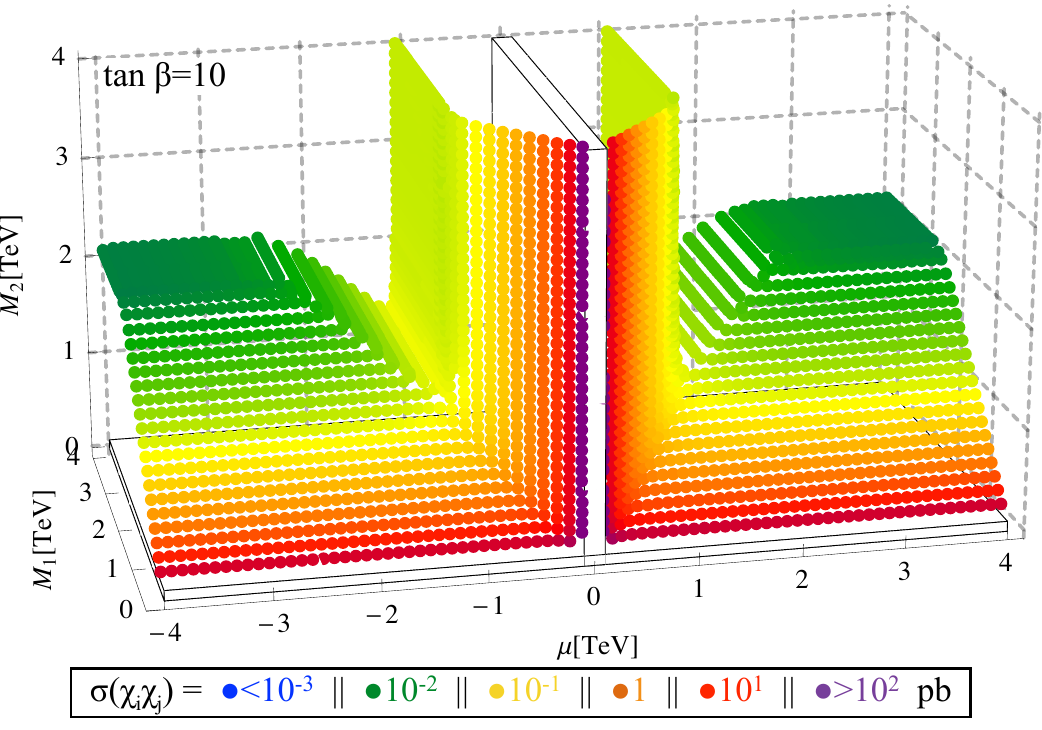} 
 \includegraphics[width=0.49\textwidth]{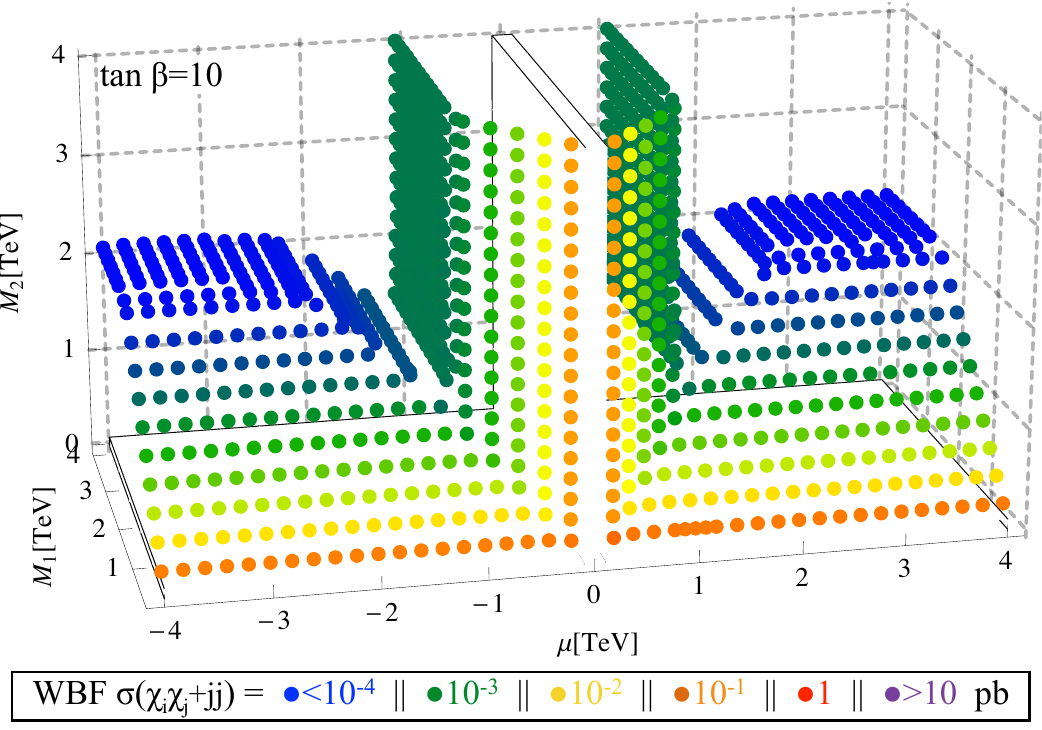} 
\caption{Relic neutralino surface with cross-sections for direct
  pair production $pp \to \chi \chi$ (left),
  and weak-boson-fusion pair production $pp \to \chi \chi jj$
  (right), at $100~\tev$ collider energy. At each point, all contributions for
  all electroweakino final states are summed. Regions ruled out by LEP
  are occluded with a white box.}
\label{fig:surface_csx}
\end{figure}

% \url{http://www3.nd.edu/~bostdiek/research_welltmp.html}
To generate the relic neutralino surface shown in
Figures~\ref{fig:surface_ann} to~\ref{fig:surface_csx}
we calculate
MSSM masses using \textsc{Suspect3}~\cite{Djouadi:2002ze} and the frozen out
relic abundance of the LSP using
\textsc{micrOmegas3}~\cite{Belanger:2013oya}. We do
not include loop corrections to the neutralino masses, which are
dominated by the scalar states, whose masses were set to $8~\tev$, including the CP-odd
Higgs~\cite{Pierce:1996zz,Fritzsche:2002bi,Oller:2003ge}. Note that
\textsc{micrOmegas3} also calculates relic abundance at leading order. For most of the 
parameter space, after fixing the values of $M_2$ and $\mu$, we vary
$M_1$ until \textsc{micrOmegas3} produces the correct relic abundance, $\Omega
h^2 \simeq 0.12$. For parameter space where the relic abundance is
attained with a decoupled bino, notably the wino-higgsino surface, we
hold $\mu$ fixed and scan over $M_2$. Note that in
Figures~\ref{fig:surface_ann} to~\ref{fig:surface_csx} the \textsc{Suspect3} and \textsc{micrOmegas3} calculations were performed with the parameters $M_1$, $M_2$ and $\mu$, defined at the decoupled scale ($8~\tev$) and $\tan \beta$ defined at $m_Z$. If $\tan \beta=10$ at $m_Z$, this will run to $\tan \beta=9.4$ at $8~\tev$. We found that if instead all parameters are defined at $m_Z$ the relic surface moves by no more than $\sim 10\%$ in $M_1$, $M_2$ and $\mu$.  

We begin our journey across the relic neutralino surface with the
relic wino. When $|\mu|$ and $M_1$ decouple, with values above $2~\tev$,
two plateaus at $M_2 = 2~\tev$ correspond to pure wino dark matter with
a mass around $2~\tev$. For $\tan \beta=10$ the features of the relic
neutralino surface are almost perfectly symmetric around $\mu=0$. The
dominant annihilation channel for the pure wino is co-annihilation
with the close-by chargino through an off-shell $W$-boson,
subsequently decaying to light-flavor quarks. \textsc{micrOmegas3}
does not include the Sommerfeld
enhancement~\cite{Hisano:2006nn,Cirelli:2007xd,Hryczuk:2010zi,Hryczuk:2011vi,Beneke:2014gja,Beneke:2014hja}
to wino annihilation, so this surface lies below the usual
value of $M_2 \simeq 2.8~\tev$. (The relic higgsino mass, on the other hand, is unaltered by Sommerfeld annihilation enhancement \cite{Ciafaloni:2013hya}). Edges of the pure wino plateau fall off, at smaller $M_1$ and $|\mu|$ respectively, to sloped bino-wino and wino-higgsino surfaces, with either $M_1 \sim M_2$ or $M_2 \sim
|\mu|$. 
On all of these surfaces co-annihilation and mixing with the
wino bolsters the annihilation of the lightest neutralino.

An amusing effect occurs on the bino-wino slope around $M_1 \sim M_2
\sim 1.7~\tev$, where the single largest annihilation channel is
$\chi^\pm \chi^\pm \to W^\pm W^\pm$. Here, the mass difference between the
lightest neutralino and the lightest chargino is so small that the
main annihilation process leading to the observed dark matter relic
density involves the chargino and not the actual LSP.

Starting from the pure wino plateaus, the mixed LSP wino-higgsino
surface terminates in a valley against a sheet of pure higgsino relic
dark matter at $|\mu| \simeq 1.1~\tev$. For such a pure higgsino,
chargino co-annihilation is again the leading dark matter annihilation
process. On the diagonal $|\mu| \sim M_1 \sim M_2$ this valley opens
onto a ridge at the intersection of the wino-higgsino, bino-wino, and
eventually bino-higgsino surfaces. This ridge contains
bino-wino-higgsino mixing with a dominant annihilation to $W^+W^-$,
both through co-annihilation with the second-lightest neutralino and
through a $t$-channel chargino with sizable couplings to the weak
gauge bosons.

On the bino-higgsino slope, large values of $|\mu|$ imply large
neutralino masses and hence an annihilation to heavy fermions
$t\bar{t}$. Usually, this annihilation process proceeds through a
heavy Higgs state in the $s$-channel. Once we decouple the heavy Higgs
states at $8~\tev$ the same role is played by the longitudinal modes of
the $Z$-boson.  Towards smaller values of $|\mu|$ the available energy
in the scattering process drops below the $t\bar{t}$ threshold and the
main annihilation goes into $W^+ W^-$ pairs, again through a
longitudinal $Z$ in the $s$-channel coupling to the higgsino
components of the lightest neutralino.

Unlike a pure wino LSP and a pure higgsino LSP the relic neutralino
surface does not feature a pure bino LSP, because in the absence of
co-annihilating scalars it would not be able to produce a relic density small enough to fit
observation. 

%---------------------------------------
%---------------------------------------

In models where the only low-energy part of the supersymmetric
spectrum are the electroweakinos, the mass difference between the
neutralino LSP and the second-lightest neutralino and lightest
chargino is crucial, not only for the dark matter (co-)annihilation
rate, but also for electroweakino decay signatures at colliders. From the
discussion of Figure~\ref{fig:surface_ann} we know that chargino co-annihilation is crucial to obtain the correct relic
density. The typical mass splitting required for co-annihilation is
$\mathcal{O}(10\%)$, or less. In Figure~\ref{fig:surface_split} we
show the absolute mass difference between the LSP and the
second-lightest neutralino (left) and the lightest chargino
(right). Over almost the entire parameter space the relative mass
splitting stays below $5\%$, except for part of the diagonal
bino-higgsino sheets. This confirms that
co-annihilation largely determines the structure of the relic neutralino
surface.

For neutralino and chargino decays the absolute mass difference (GeV) is
more relevant than the relative (percentage) mass difference, because it determines
if a $1 \to 2$ particle decay can occur through on-shell $W^\pm$ or $Z$-boson. If on-shell decays are forbidden, decays must proceed through an off-shell weak gauge boson or a loop-induced decay.

For the mass difference between the lightest two neutralinos, we see
that some of the pure wino regime will allow for an on-shell decay $\chi_2^0
\to \chi_1^0 Z$. This means that collider searches for pure winos can use
leptonic $Z$ decays, including a same-flavor opposite-sign $Z$-mass
constraint~\cite{Low:2014cba,Cirelli:2014dsa,Acharya:2014pua}. This
makes one $100~\tev$ hadron collider search strategy for pure wino dark matter
straightforward, including triggering and detector effects\footnote{In this situation, dark matter constraints alone do not give the mass of $\chi^0_2$. The mass of $\chi^0_2$ will set the rate for traditional tri-lepton search channel $pp \to \chi^{\pm}_1 \chi^0_2$.}. However,
anywhere else on the relic neutralino surface the neutralino mass difference drops below $80~\gev$ so $\chi^0_2 \to \chi^0_1$ decay via a $Z$-boson
has to occur off-shell. This includes the pure higgsino sheets, as well
as the the bino-wino, the bino-higgsino, and the wino-higgsino mixing slopes. This small inter-state splitting is one of the reasons that collider searches for electroweakino dark matter are challenging. We will
address this challenge in Section~\ref{sec:resolution}.

For much of the relic neutralino surface, the NLSP is the
lightest chargino. Its mass difference to the dark matter agent rarely
exceeds $30~\gev$, with the exception of lighter bino-higgsino dark
matter. For the collider signatures, this means that both leptons and
neutrinos from off-shell $W$ decays will typically have
small transverse momenta. 

%---------------------------------------
%---------------------------------------

To determine the dark matter complementarity reach of a $100~\tev$ collider, 
it is useful to consider what regions of the
relic neutralino surface will be accessible to current and future
dark matter searches. Bounds from direct detection and neutrino telescope experiments \cite{Xiao:2014xyn,Agnese:2014aze,Akerib:2013tjd,Aprile:2013doa,Aprile:2012nq,Behnke:2010xt,Felizardo:2011uw,Archambault:2012pm,Aartsen:2012kia} constrain TeV-mass dark matter to have a spin-independent cross-section less than $10^{-44}~\rm{cm^2}$ and spin-dependent cross-section less than $10^{-39}~\rm{cm^2}$. After calculating the cross-section for
the lightest neutralino scattering off protons and neutrons in
\textsc{micrOmegas3}, in Figure~\ref{fig:surface_dd} we show the spin
dependent and spin independent nuclear cross sections relevant to
direct detection across the relic neutralino surface. The leading order scattering cross-sections given in Figure~\ref{fig:surface_dd} indicate that much of the relic neutralino surface
is beyond the detection capability of current or planned direct
detection experiments.  While bino-higgsino dark matter and the bottom of the wino-higgsino valleys shown in Figure~\ref{fig:surface_dd} are presently already excluded by the LUX experiment~\cite{Akerib:2013tjd} assuming nominal values for astrophysical and nuclear inputs,
the bino-wino portion of the relic surface presents a particular challenge. As $|\mu|$ increases, the higgsino
component of the LSP and the corresponding LSP coupling to the Higgs
both decrease, causing a major diminution of both spin-dependent and
independent nucleon cross-sections. 
Future experiments will also have to contend with a solar neutrino background that will be relevant for spin-independent cross-sections less than $10^{-48}~\rm{cm^2}$, as detailed in the introduction.  

%---------------------------------------
%---------------------------------------

To begin addressing the challenge of discovering electroweakino dark
matter at a future collider, in Figure~\ref{fig:surface_csx} we show
the $100~\tev$ hadron collider production rates for electroweakino
pair production, in the direct $2 \to 2$ production process (left) and
in the weak boson fusion jet-associated production
(right). The direct production rates are computed with
\textsc{Prospino2}~\cite{Beenakker:1999xh} while for the WBF
process we rely on \textsc{Madgraph}~\cite{Cho:2006sx}. 
All electroweakino pair combinations are summed. The rates range from $100~\pb$ for
low masses down to $0.01~\pb$.  Except along the bino-wino-higgsino ridge, 
the cross sections will be dominated by combinations
of the lightest three states, $\chi_{1,2}^0$ and $\chi_1^\pm$. While
the neutralino coupling to a $Z$-boson is driven by the higgsino
content, the chargino coupling to photons and to the $Z$-boson
includes the wino as well as the higgsino fraction. The mixed
neutralino--chargino coupling to a $W$-boson is diagonal in the
gaugino and higgsino fractions, respectively. In the usual bino-LSP
scenarios probed at the LHC the leading production processes are
$\sigma(\chi_2^0 \chi_1^\pm) > \sigma(\chi_1^+ \chi_1^-)$. The
neutralino rates $\sigma(\chi_2^0 \chi_2^0) \sim \sigma(\chi_1^0
\chi_2^0)$ are typically smaller~\cite{Beenakker:1999xh}.

If the neutralino LSP and the chargino NLSP are pure winos, as is the
case on the wino LSP plateaus, direct chargino pair production and
$\chi_1^0 \chi_1^\pm$ production will have an un-suppressed rate. Its
size is determined by the masses of the lightest neutralinos and
charginos. Pure higgsino pairs couple to photons, $W$-bosons, and
$Z$-bosons in the $s$-channel, which in combination with the lower
mass scale leads to larger cross sections on the higgsino
sheets. Adding a bino fraction to the LSP will not lead directly to a
significant increase of the direct production rate, because the bino
fraction does not couple to $W$- or $Z$-bosons. However, on the relic
neutralino surface the bino fraction reduces the dark matter
annihilation cross-section and thereby drives the required mass scales
lower. This rapidly increases the $100~\tev$ hadron collider cross sections for direct
electroweakino pair production to $0.1~\pb$ for an LSP mass around
$500~\gev$.

The difference between direct production and weak-boson-fusion
production of electroweakinos~\cite{Cho:2006sx}
is that the latter can be driven exclusively by the
coupling to electroweak gauge bosons. 
Unfortunately, we seem
to observe no improvements to speak of, compared to the direct
production mode. Instead, the WBF production rates are roughly an
order of magnitude below the direct production rate. In
the next section we will find that the boosted kinematics 
are the primary factor that extends the
discovery potential of a $100~\tev$ collider across the relic
neutralino surface. We do not pursue the WBF production further in this paper; however, as electroweakinos produced via WBF tend to be boosted, these events could 
help to extend the $100~\tev$ collider discovery potential using the strategy 
shown in the next section.

%%%%%%%%%%%%%%%%%%%%Section Resolution%%%%%%%%%%%%%%%%%%%%%%%%%
\section{Almost degenerate dark matter searches}
\label{sec:resolution}
%%%%%%%%%%%%%%%%%%%%Section Resolution%%%%%%%%%%%%%%%%%%%%%%%%%

Having mapped out the regions of electroweakino parameter space which
predicts the correct relic abundance, we turn to collider studies and
focus our attention on some collider-salient features of the relic
density surface. First, the splitting between LSP and its
electroweakino cohorts is small. From Figure~\ref{fig:surface_split}
we know that the neutralino--neutralino splitting is $\Delta
m_{\chi_2^0-\chi_1^0} < m_Z$ for all electroweakino admixtures with
the sole exception of the pure wino plateau. The chargino--neutralino
splitting stays below $\Delta m_{\chi_1^\pm-\chi_1^0} < m_W$ over the
entire surface.  A neutralino mass splitting below the $Z$-mass
complicates hadron collider searches, because it spoils one of the
most effective top-background rejection cuts in tri-lepton
searches \cite{Chatrchyan:2013dsa,Chatrchyan:2014aea,Aad:2014nua,Aad:2014yka,Calibbi:2013poa,Calibbi:2014lga}. The compressed nature of the relic electroweakino sector with its
soft decay leptons challenges vanilla electroweakino pair production
$pp \to \chi_i \chi_j$; issues arise with triggering, as well as
background rejection. In the analysis following, we explore boosted electroweakinos recoiling
against hard jets as a unique opportunity at a $100~\tev$
collider. After accounting for the decay of any heavier
electroweakinos, boosted electroweakino final states have the form $pp
\to j \met X$, where $X$ is some combination of leptons and photons.

The hard jet in the event serves two purposes: first, it can be
triggered on and, in turn, allows the cuts on photons or leptons to be
relaxed. This is crucial when we want to be sensitive to decays of the
heavier electroweakinos with small mass splittings. Second, it impels
the electroweakinos in the opposite direction, which leads to large
missing transverse momentum. We will see that the combination of large
$\met$ plus soft leptons or photons is especially powerful at a
$100~\tev$ collider. 

While the basic kinematics of boosted electroweakino signals are
dictated by the hard jet in the event, the details depend on what soft
particles are present. As these soft particles come from heavier
electroweakino decays, their identity is dictated by what combination
of chargino or neutralino states are produced.  This, in turn, is
dictated by the wino/bino/higgsino content of the electroweakinos. In
the following subsections we present two possible scenarios,
\begin{alignat}{5} 
pp &\to \ell^\pm \gamma \, j \, \met \notag \\
pp &\to \ell^+ \ell^- \, j\,  \met \; ,
\label{eq:processes_gen}
\end{alignat}
which will be particularly effective in finding compressed
electroweakino decays at a $100~\tev$ hadron collider. In the
combination of the two channels we will see that the $100~\tev$ hadron
collider can resolve mass splittings below $5~\gev$ between TeV-mass
electroweakino states.

To establish the coverage of the relic neutralino surface, we proceed
in three steps: first, in Section~\ref{sec:well-tempered} we will
apply the lepton--photon analysis to the relic neutralino surface,
specifically to bino-wino dark matter. In
Section~\ref{sec:sensitivity} we will examine this analysis strategy
more generally, considering neutralino mass splittings beyond those attained across the relic surface.  This has implications for dark
matter outside of the MSSM, as we will determine the $100~\tev$
collider sensitivity to $1-25~\gev$ mass splitting for triplet-singlet
$SU(2)$ states. In Section~\ref{sec:leptons} we will show the coverage
of the relic neutralino surface at slightly larger mass splittings of
$5-50~\gev$ using the dilepton decay mode~\cite{Giudice:2010wb,Gori:2013ala,Schwaller:2013baa,Han:2014kaa,Baer:2014kya,Han:2014sya}. Soft lepton
studies of non-relic electroweakinos at $100~\tev$ have been addressed previously in
the literature~\cite{Low:2014cba}. Our study considers a wider range of
inter-electroweakino splittings than~\cite{Low:2014cba} and applies the constraint that the LSP lies on the relic neutralino surface.

%%%%%%%%%%%%%%%%%%%%%%%%%%%%%%%%%%%%%%%%%%%%%
\subsection{Surfing the relic surface with a photon and lepton}
\label{sec:well-tempered}
%%%%%%%%%%%%%%%%%%%%%%%%%%%%%%%%%%%%%%%%%%%%%

%---------------------------------------
\begin{figure}[t]
\includegraphics[width=.49\textwidth]{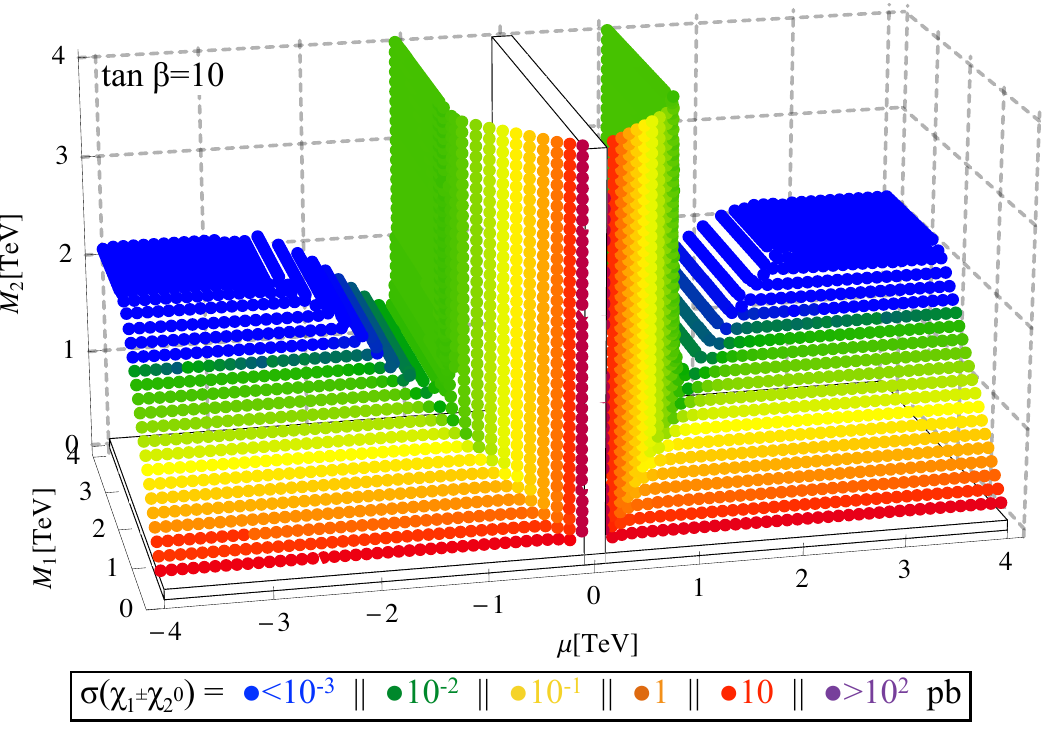}
\includegraphics[width=.49\textwidth]{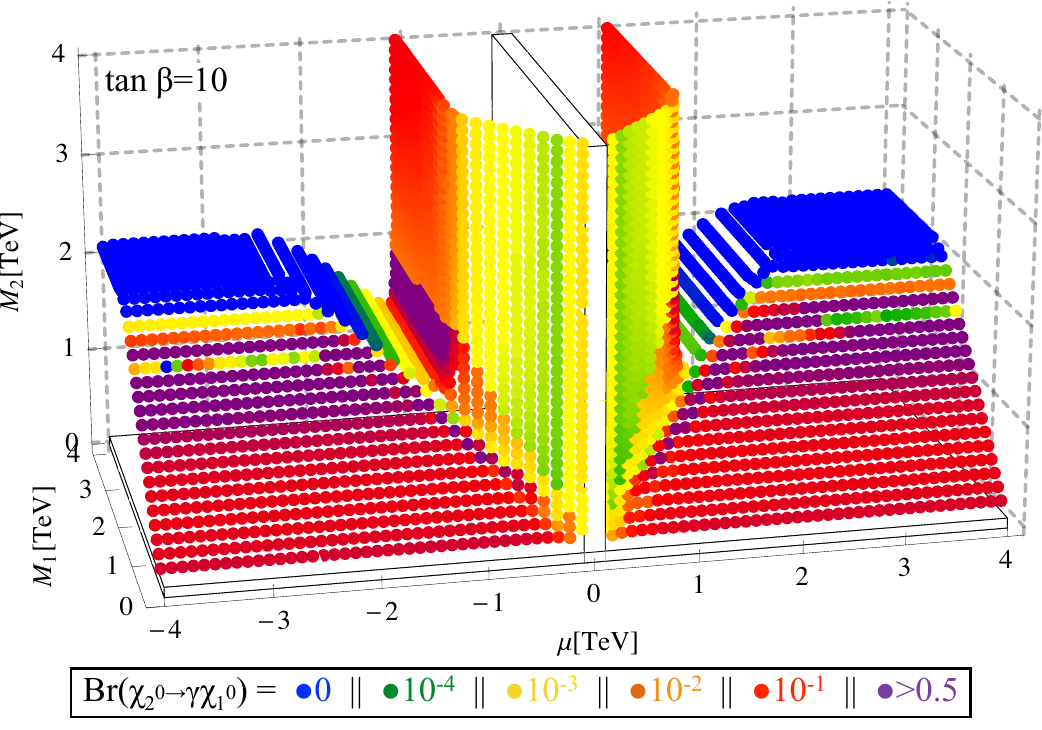}
\caption{Relic neutralino surface with the $\chi_2^0 \chi_1^\pm$
  production cross-section (left) and the $\chi_2^0 \to \chi_1^0
  \gamma$ branching ratio (right).}
\label{fig:surface_sigmabr1}
\end{figure}
%---------------------------------------

The first channel we examine for discovering the MSSM dark matter
spectrum is the lepton--photon signature shown in
Eq.\eqref{eq:processes_gen}.  The soft photon comes from radiative
decays $\chi_i^0 \to \gamma \chi_1^0$. As inter-neutralino splittings
become small, which they do across the bino-wino surface and the
bottom of the wino-higgsino surface shown in
Figure~\ref{fig:surface_split}, the radiative decay to photons becomes
competitive with increasingly off-shell decays through $Z$-bosons. In
Figure~\ref{fig:surface_sigmabr1}, we display the branching ratio of
the second lightest neutralino to a photon and the lightest
neutralino. It is sizable across the bino-wino surface, for a narrow
piece of the wino-higgsino surface, and also across the pure higgsino
plane, where there is a small splitting between neutral higgsino
states.  These regions will be difficult to reach with future direct
detection searches, as we learned from Figure~\ref{fig:surface_dd}.

%---------------------------------------
\begin{figure}[t]
\begin{center}
\includegraphics[width=.95\textwidth]{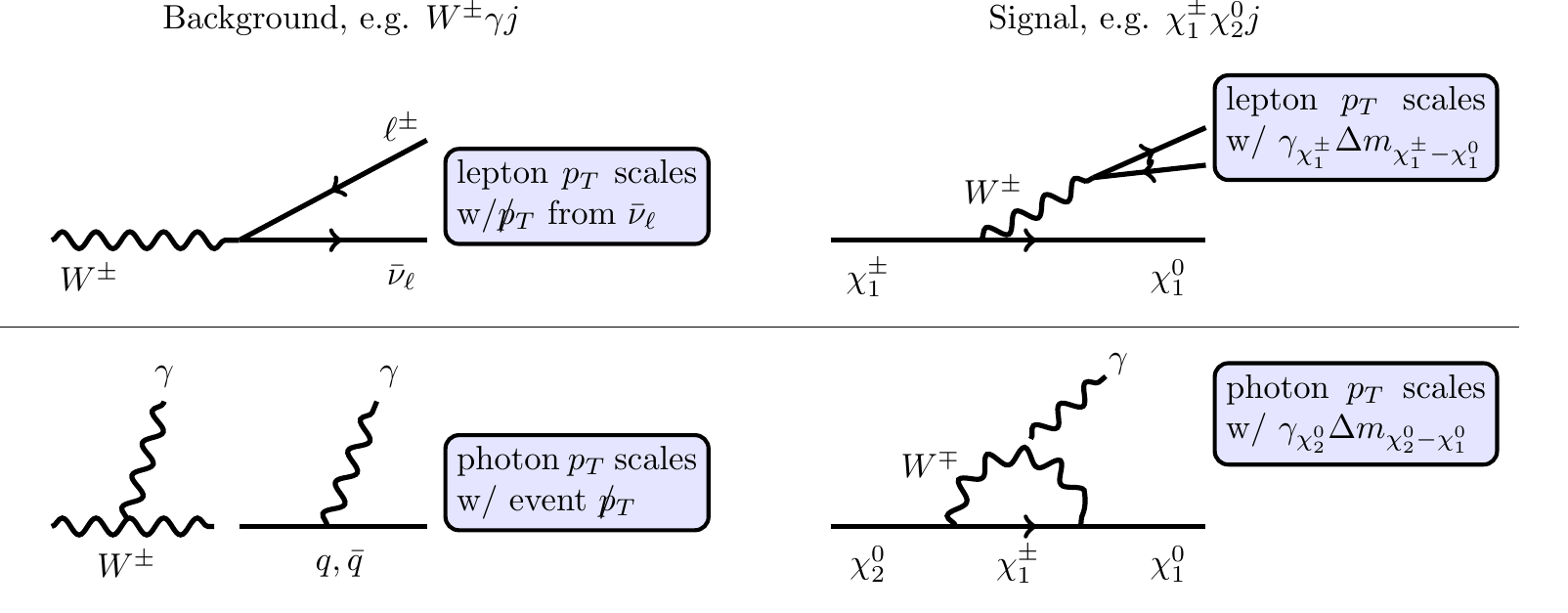}
\end{center}
\caption{Illustration of how $p_T^\text{max}$ cuts on the lepton or
  photon reduce the $W\gamma$ background. These cuts will
  gain in efficacy as missing transverse momentum becomes much larger
  than inter-electroweakino mass splittings, as will occur at a
  $100~\tev$ collider, because the electroweakino Lorentz boosts 
  ($\gamma_{\chi_2^0}$ and $\gamma_{\chi_1^\pm}$) and mass splittings
  typically produce smaller lepton and photon transverse momenta.}
\label{fig:boostelectro}
\end{figure}
%---------------------------------------

The electroweakino combination that most easily generates a final state with a lepton and photon is $\chi^{\pm}_1\chi^0_2$. To see how prevalent this electroweakino combination is as we traverse the relic neutrino surface, we plot the $pp \to \chi^{\pm}_1\chi^0_2$ cross section at $100\,\tev$ in the left panel of Figure~\ref{fig:surface_sigmabr1}. The rate is large, ($\gtrsim 10\,\fb$) for most of the relic neutralino surface. In particular, the $\chi^{\pm}_1\chi^0_2$ cross section is sizable even in the bino-wino regions where the spin-independent cross are so small they sit below the solar neutrino background cross-section (see Figure~\ref{fig:surface_dd}). 

The cross section we are really interested in is not $pp \to \chi_1^{\pm}\,\chi^0_2$, but electroweakinos produced in association with hard initial state radiation, $pp \to \chi_1^{\pm}\,\chi^0_2+j$. Accounting for the extra radiation, the cross sections shown in Fig.~\ref{fig:surface_sigmabr1} need to be adjusted, however this adjustment is a function of the $p_T$ of the radiated jet and will be the same for all electroweakino processes. The final state
\begin{equation}
pp \to \chi_2^0 \chi_1^\pm \, j 
   \to \left( \gamma \chi_1^0 \right) \, 
       \left( \ell^\pm \nu \chi_1^0 \right) \, j
\label{eq:process1}
\end{equation}
is effective because the background can be reduced by requiring a
soft photon and a lepton in association with a large amount of missing
transverse momentum and a hard jet.  The underlying decay processes are
illustrated in Figure~\ref{fig:boostelectro}. The dominant background
to the neutralino-chargino signature is $pp \to W_\ell^\pm \gamma j$.
A missing transverse momentum cut in the TeV range makes direct use of the
increased collider energy of $100~\tev$. For the background, the
$W$-boson has to be strongly boosted itself, giving $p_{T,\ell} \sim
p_{T,\nu} \sim \met$. The background photon will sometimes inherit a
significant amount of transverse momentum from recoiling against a
very hard jet and a very hard $W$-boson.  In contrast, electroweakino
decays produce large missing transverse momentum through the boosted
pair production process with two un-balanced LSPs.  The lepton
momentum will be set by the inter-electroweakino mass splitting,
$p_{T,\ell} \propto \gamma_{\chi} \Delta m_{\chi-\chi_1^0}/2$, where
$\gamma_{\chi}$ is the boost factor of the heavy decaying electroweakino, as
illustrated in Figure~\ref{fig:boostelectro}. Altogether,
this allows for efficacious electroweakino searches at a $100~\tev$
hadron collider sensitive to $p_T = 5-50~\gev$ photons and
leptons in events triggered by large $\met$.

%---------------------------------------
\begin{figure}[t]
\includegraphics[width=.99\textwidth]{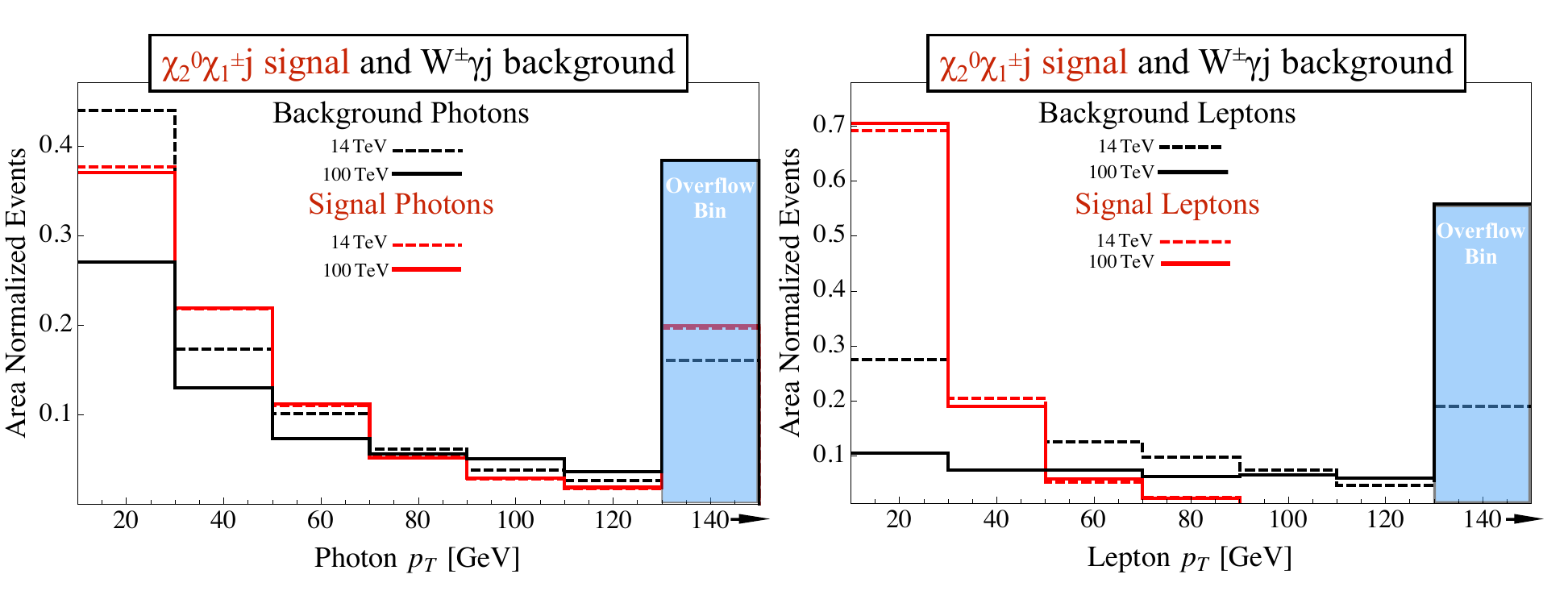}
\caption{Photon and lepton transverse momenta for the $\chi_2^0
  \chi_1^\pm j$ signal and the $W \gamma j$ background at $14~\tev$
  and $100~\tev$. We assume a bino--wino mass spectrum $m_{\chi_2^0} =
  m_{\chi_1^\pm} = 200~\gev$ with a mass splitting $\Delta
  m_{\chi_2^0-\chi_1^0} =10~\gev$. At this level we require at least
  one jet with $p_{T,j} > 100~\gev$ for the $14~\tev$ collider and
  $p_{T,j} > 600~\gev$ for $100~\tev$. Note that the tails of these distributions,
  here shown collected in an overflow bin, fall off out to $p_T \sim \tev.$}
\label{fig:photsandleps}
\end{figure}
%---------------------------------------
   
To determine how well the hard missing transverse momentum cut together with lepton
and photon cuts discriminate the electroweakino signal from the
$W\gamma j$ background, we generate tree-level
signal and background events in
\textsc{Madgraph5}~\cite{Alwall:2014hca}\footnote{We used the default \textsc{Madgraph5} parton density functions, factorization and renormalization schemes for all simulated events.} combined with
\textsc{Pythia6.4}~\cite{Sjostrand:2006za} and the anti-$k_T$ jet
algorithm~\cite{Cacciari:2008gp,Cacciari:2011ma} for clustering partons into jets,
with $R=0.5$.  We simulate the detector acceptance using
\textsc{Delphes3}~\cite{deFavereau:2013fsa}, with the Snowmass
detector card~\cite{Anderson:2013kxz}. For generator-level cuts 
we require one jet with $p_{T,j} > 600~\gev$ and a minimum
missing transverse momentum $\met > 1.5~\tev$.
Our results only rely on the leading order $pp \to
\chi_2^0 \chi_1^\pm j$ cross-sections, which will be increased by NLO
contributions~\cite{Beenakker:1999xh,Cullen:2012eh,Fuks:2012qx},
threshold and transverse momentum
resummation~\cite{Debove:2009ia,Debove:2011xj}, and weak boson
fusion~\cite{Cho:2006sx}. 

To illustrate why this analysis works with maximum cuts on lepton and
photon momenta, we show the lepton and photon transverse momentum
distributions for a $m_\chi = 200 ~\gev$ bino-wino with a 10 GeV
inter-neutralino mass splitting in Figure~\ref{fig:photsandleps}. For
the LHC with $14~\tev$ we find many $W \gamma j$ background events with a soft
lepton and a soft photon. As the collider energy and the cuts on the
hard jet and the missing transverse momentum increase, the background
lepton and photon become harder. Both of them show a
correlation with the missing transverse momentum cut of $\met > 1.5~\tev$,
making it easier to remove this background with a maximum photon and
lepton $p_T$ requirement.\\

We now proceed to the analysis of the bino-wino portion of the relic
neutralino surface. To probe this parameter regime we decouple the
higgsino fraction at $|\mu| = 4~\tev$. Adjusting $M_1$ and $M_2$
allows us to follow the line with the correct relic density. As the mass splitting varies as we traverse the surface, one would ideally optimize the cuts at each $M_1, M_2$ point to maximize the efficiency. Here, for simplicity, we work with only two sets of cuts; one set for $M_1$ below $\sim 900\,\gev$, where the LSP is more bino-like, and one set for $M_1 > 900\,\gev$ where the LSP is more wino.

\begin{alignat}{5}
p_{T,\ell} &=
\begin{cases}
 [5,80]~\gev \quad (M_1 < 900~\gev) \\
 [5,40]~\gev \quad (M_1 \ge 900~\gev) 
 \end{cases}
\qqquad & |\eta_\ell| &< 2.5 \notag \\
p_{T,\gamma} &=
\begin{cases}
 [5,80]~\gev \quad (M_1 < 900~\gev) \\
 [5,60]~\gev \quad (M_1 \ge 900~\gev) \\
 \end{cases}
\qqquad & |\eta_\gamma| &<2.5 \qqquad \qqquad & \Delta R_{\ell-\gamma} &> 0.5 \notag \\
p_{T,j} &> 1~\tev \qqquad & |\eta_j| &<2.5 \notag \\
\met &> 1.5~\tev
\label{eq:temperedcuts}
\end{alignat}
To reject hadronic backgrounds, the cuts require no more than two
jets with $p_T > 300~\gev$. The jet cut given in
Eq.\eqref{eq:temperedcuts} is really only relevant for triggering; lowering
it to $p_{T,j} > 300~\gev$ does not affect the collider reach.  An
improved lepton-photon analysis would vary the jet and missing transverse momentum cuts as a function of the electroweakino
masses.

The lower cut on the soft lepton and photon $p_T$ is an optimistic assumption on the detector performance, however the utility of the low threshold motivates taking these values seriously.
 The Snowmass \textsc{Delphes3} card~\cite{Anderson:2013kxz} assumes zero efficiency for leptons with less than $10\,\gev$ of energy. Therefore, for the purposes of this study we modify the card, matching the
sensitivity for $5-10~\gev$ photons and leptons to the value at $10\,\gev$. However, even without sensitivity to 
 photons and leptons $< 10\,\gev$, the required luminosity for detecting all points 
shown remains less than ten inverse attobarns.

Like any photon-based analysis this electroweakino search relies on an
efficient rejection of fake-photons from jets. Focusing on the $M_1
\ge 900\,\gev$ analysis, we generate hard $W+\text{jets}$ events,
again with \textsc{Madgraph5}, \textsc{Pythia6.4}, and
\textsc{Delphes3}. Lowering the \textsc{Delphes3} jet threshold to
$5\,\gev$ and treating every radiated jet with $p_{T,j} < 60~\gev$ as
a possible fake photon we determine the required fake photon rejection
rate.  We find a rejection rate around $1/125$ for $p_{T,\gamma} >
5~\gev$ is needed to safely suppress the fake background. This
estimate is only approximate as it ignores any kinematic dependence in
the fake rate and assumes our tool settings (in \textsc{Madgraph5} and
\textsc{Pythia6.4}) accurately describe the $100~\tev$ collision
environment. However, the fake rate we need to suppress
$W+\text{jets}$ events is orders of magnitude more conservative than
what is currently achievable at the LHC~\cite{ATL-PHYS-PUB-2011-007}.

%---------------------------------------
\begin{figure}[t]
\begin{center}
\begin{tabular}{c}
\includegraphics[width=.6\textwidth]{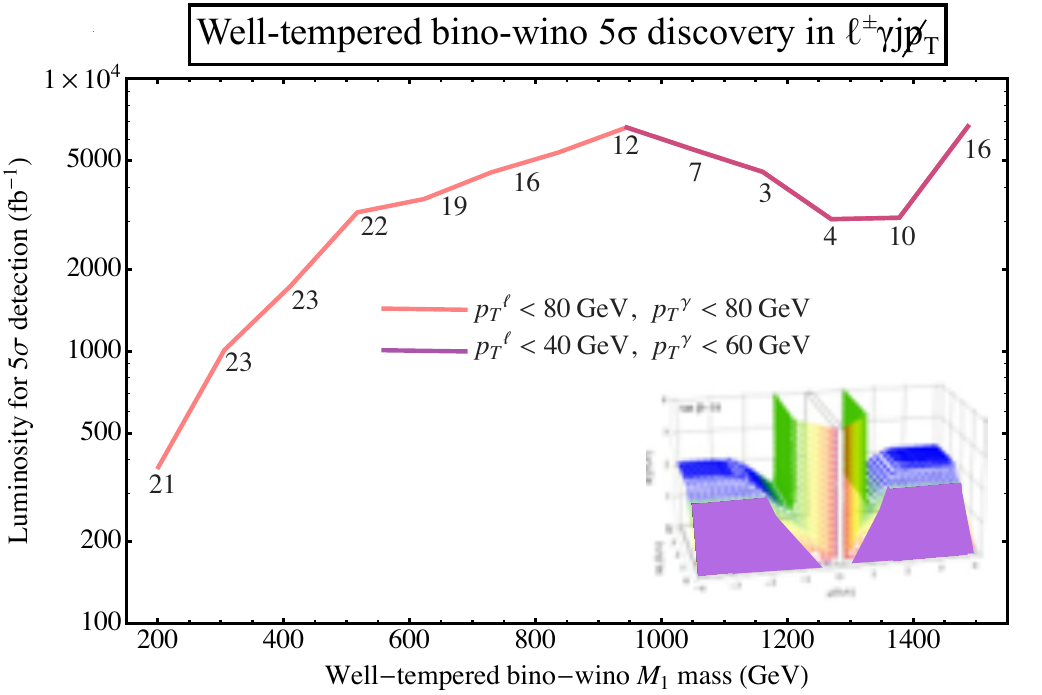}
\end{tabular}
\end{center}
\caption{Well-tempered bino--wino mass reach at a $100~\tev$ proton
  collider in the $\ell \gamma j \met$ final state for $M_1 =
  0.2-1.5~\tev$. The mass splitting in GeV between the $\chi_2^0$ and the LSP is
  indicated at each point on the curve. Note that the mass splitting between
  $\chi_1^\pm$ and the LSP will be $\sim 0.5~ \gev$ larger than that of $\chi_2^0$, 
  because $\chi_2^0$ is mostly wino.}
\label{fig:welltemplum}
\end{figure}
%---------------------------------------

In Figure~\ref{fig:welltemplum} we show the luminosity required to
discover bino-winos on the relic neutralino surface at a $100~\tev$
hadron collider. Here we take the simplest definition of significance as
$S/\sqrt{B}$, and note that all points in Figure 8 have $S/B > 1/5$ and more than fifty
signal events expected. With around seven inverse attobarns of
integrated luminosity, the relic bino-wino surface will be detected
with $5 \sigma$ significance for LSP masses up to $1.5~\tev$. We checked that assuming 10\% systematic errors for the signal and 2\% for the background, still permits $5 \sigma$ significance detection with 10 ab$^{-1}$ luminosity. Above an LSP mass of $1.5~\tev$, parameter space transitions to the pure wino surface, as indicated by the
rapidly increasing NLSP-LSP mass splitting. Although we have carried out this analysis for $\mu=4\,\tev$ we expect the results to apply over much of the bino-wino surface, as indicated by the shaded region in Figure~\ref{fig:welltemplum}, as the relevant properties do not vary much as $\mu$ is lowered.

The reach of our photon signature can be compared to the parameter
coverage through charged tracks~\cite{Low:2014cba}. While charged
track searches should be able to detect the wino component over portions of the
bino-wino relic surface at a $100~\tev$ collider, underlying
assumptions about the superpartner mass spectrum are crucial.  Custodial symmetry
breaking in the Standard Model induces a $160~\mev$ splitting between
the neutral and charged wino states~\cite{Ibe:2012sx}. However, such a
small chargino-LSP splitting is not guaranteed; it assumes that both
$|\mu|$ and every other $SU(2)_L$ MSSM mass is decoupled above
$10~\tev$. For example a higgsino mass of $|\mu| \sim 1.5~\tev$ will
result in an additional $200~\mev$ tree-level splitting between the
charged and neutral wino states of an $800~\gev$ bino-wino. The electroweak NLO corrections to
the second-lightest neutralino in the presence of light scalars easily
exceed $1~\gev$, while keeping the LSP mass
constant~\cite{Pierce:1996zz,Fritzsche:2002bi,Oller:2003ge}. Decoupling
the scalars will reduce the typical size of these electroweak
corrections, but the corrections from the electroweakino sector itself
will not necessarily drop below $1~\gev$. This is why a second channel
covering electroweakino dark matter with small mass differences
significantly adds to the case for a $100~\tev$ collider.

%%%%%%%%%%%%%%%%%%%%%%%%%%%%%%%%%%%%%%%%%%%%%
\subsection{Probing small neutralino mass splittings}
\label{sec:sensitivity}
%%%%%%%%%%%%%%%%%%%%%%%%%%%%%%%%%%%%%%%%%%%%%

Going beyond the discovery of relic bino-winos, we now apply the same
photon plus lepton search to a broader range of bino-wino mass splittings and masses, but without imposing the constraint of viable relic abundance. To determine how small a mass difference between
weakly charged particles a $100~\tev$ collider can resolve, we assume
similar cuts and procedures as the preceding section.  Throughout this
section as throughout the last, we decouple the higgsino content at
$\mu = 4~\tev$.

%---------------------------------------
\begin{figure}[t]
\includegraphics[width=.99\textwidth]{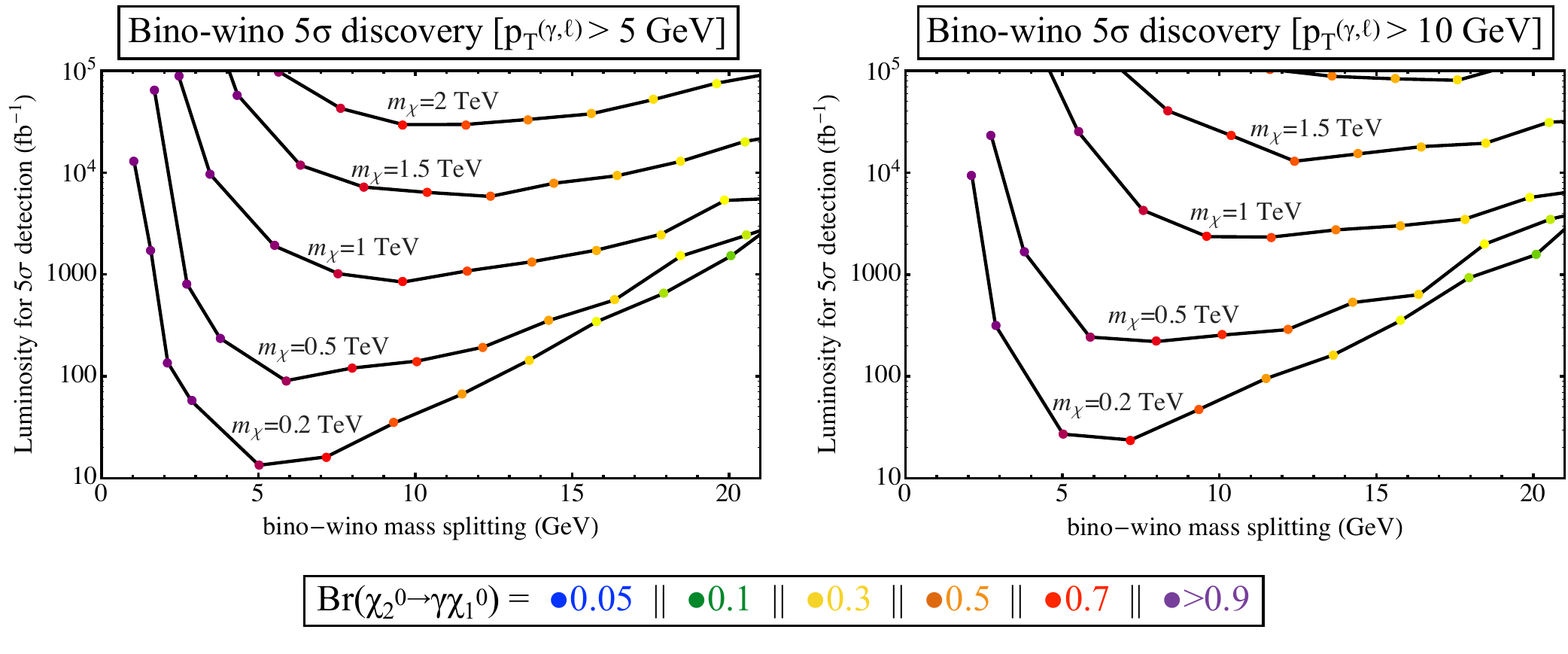} 
\caption{$100~\tev$ hadron collider mass reach for a range of
  bino--wino mass splittings. The LSP mass is indicated on each
  curve. The cuts applied are shown in Eq.\eqref{eq:bwcuts} and the
  corresponding text.  The branching fraction for the decay $\chi_2^0
  \to \chi_1^0 \gamma$ is indicated at each significance point.}
\label{fig:splitlum}
\end{figure}   
%---------------------------------------

We compute the luminosity that will be required
for a detection of compressed bino-winos as a function of the LSP mass and the mass splitting to the
second-lightest neutral state (the splitting between the LSP and the lightest chargino is similar). These target luminosities again
assume $W \gamma j$ as the dominant background. We cover mass
splittings of $1-25~\gev$ in two steps. First, for
neutralino mass splittings $2~\gev$ or greater, we require exactly one lepton
($e$ or $\mu$), one photon, and at least one jet, with
\begin{alignat}{5}
p_{T,\ell} & = [10,40]~\gev \qqquad & |\eta_\ell| &< 2.5 \notag \\
p_{T,\gamma} &= [10,60]~\gev \qqquad & |\eta_\gamma| &<2.5 \qqquad \qqquad & \Delta R_{\ell-\gamma} &> 0.5 \notag \\
p_{T,j} &> 1~\tev \qqquad & |\eta_j| &<2.5 \notag \\
\met &> 
\begin{cases} 
  1.5~\tev \quad (m_{\chi_2^0} = 200~\gev) \\
  2.0~\tev \quad (m_{\chi_2^0} > 200~\gev) \; .
\end{cases}
\label{eq:bwcuts}
\end{alignat}
For mass splittings between $1~\gev$ and $2~\gev$ and a
low LSP mass of $m_{\chi_1^0} = 200~\gev$, the lepton and photon
acceptance cuts in Eq.\eqref{eq:bwcuts} are not efficient enough. For
such small splittings, we lower the lepton and photon cuts
to $p_{T,\ell} =[3,10]~\gev$ and $p_{T,\gamma} = [5,10]~\gev$; we consider these acceptance cuts a benchmark value and expect they could be optimized further. From the right panel of
Figure~\ref{fig:surface_split} we see that such a small leading--order
mass splitting indeed occurs in the pure wino and higgsino regions.
As discussed above, it needs to be seen how large quantum corrections
to this mass splittings can
be~\cite{Pierce:1996zz,Fritzsche:2002bi,Oller:2003ge}. 

In Figure~\ref{fig:splitlum} we show the luminosity which is required
for a $5 \sigma$ discovery of associated neutralino-chargino
production with the photon decay signature shown in
Eq.\eqref{eq:process1}. Again, we define significance simply as $S/\sqrt{B}$ and note that
$S/B > 1/4$ for all points shown, except for $m_{\chi_1^0} \sim 2 ~\tev$ where $S/B > 1/20$. 
The two relevant parameters are the mass difference between the lightest two neutralinos, determining the
kinematics, and the mass scale of the two produced particles,
determining the production cross-section. For this first result we do
not require the neutralino LSP to reproduce the observed dark matter
relic density. We show the conservative detector acceptance cuts from
Eq.\eqref{eq:bwcuts}, assuming sensitivity to $p_T > 10~\gev$ photons and leptons,
as well as results for an improved lower cut for the lepton and photon momenta of $p_T > 5~\gev$. 

In this general analysis we see that the properties that make a
$100~\tev$ collider ideal for dark matter mass spectroscopy are simple
but powerful: the signature including a very hard jet and sizeable
missing transverse momentum as well as a lepton and a photon will feature extremely
boosted electroweak backgrounds.  In contrast, heavy electroweakino pairs
will be produced closer to threshold, and the decays of
lighter neutralinos and charginos will be so boosted that the
neutralino mass splittings in the detector's rest frame will 
be resolvable as otherwise un-detectably soft leptons and
photons. While bino--winos of up to $150~\gev$ in mass and with $> 10~\gev$ inter-state splittings can be probed at the  LHC~\cite{Bramante:2014dza,Han:2014xoa}, a $100~\tev$ collider can resolve mass splittings as small as a GeV and is sensitive to much heavier
neutralinos. 

%%%%%%%%%%%%%%%%%%%%%%%%%%%%%%%%%%%%%%%%%%%%%
\subsection{Cruising the relic surface with soft leptons}
\label{sec:leptons}
%%%%%%%%%%%%%%%%%%%%%%%%%%%%%%%%%%%%%%%%%%%%%

Moving from few-GeV mass splittings to mass splittings between
$5~\gev$ and $50~\gev$ we now establish the impact of the second,
dilepton signature in Eq.\eqref{eq:processes_gen}.  Our analysis
largely follows its LHC counterpart~\cite{Han:2014kaa} and aims for
the gap between lepton--photon or mono-jet analyses for very small
mass differences, and the tri-lepton search which is most successful
for mass splittings above $50~\gev$. On the relic neutralino surface,
the lepton-photon channel described in Section~\ref{sec:well-tempered}
and the soft-lepton signature will complement each other on the
bino-wino slope. In addition, soft dileptons will be the leading
search strategy on the bino--higgsino sheet. Some relevant signal
processes contributing to the soft dilepton signature, ordered by
typical size, are
\begin{alignat}{5} 
pp &\to \chi_1^+ \chi_1^- \, j 
   \to \left( \ell^+ \nu \chi_1^0 \right) \, 
       \left( \ell^- \bar{\nu} \chi_1^0 \right) \, j \notag \\
pp &\to \chi_2^0 \chi_1^\pm \, j 
   \to \left( \ell^\pm \nu jj \chi_1^0 \right) \, 
       \left( \ell^\pm \nu \chi_1^0 \right)  \, j \notag \\
pp &\to \chi_2^0 \chi_1^\pm \, j 
   \to \left( \ell^+ \ell^- \nu \bar{\nu} \chi_1^0 \right) \, 
      \left( jj \chi_1^0 \right)  \, j \notag \\
pp &\to \chi_2^0 \chi_1^\pm \, j 
   \to \left( \ell^+ \ell^- \chi_1^0 \right) \, 
       \left( jj \chi_1^0 \right) \, j \notag \\
pp &\to \chi_2^0 \chi_1^\pm \, j 
   \to \left( \ell^+ \ell^- \chi_1^0 \right) \, 
       \left( \ell^\pm \nu \chi_1^0 \right) \, j \notag \\
pp &\to \chi_2^0 \chi_1^\pm \, j 
   \to \left( \ell^+ \ell^- \nu \bar{\nu} \chi_1^0 \right) \, 
       \left( \ell^\pm \nu \chi_1^0 \right) \, j \notag \\
pp &\to \chi_2^0 \chi_1^0 \, j 
   \to \left( \ell^+ \ell^- \chi_1^0 \right) \, 
       \chi_1^0  \, j \; .
\label{eq:process2}
\end{alignat}
Channels involving a $\chi_2^0$ pair are not
considered since the underlying production cross section is negligible.
When the spectrum is compressed, as we have shown it is for almost the
entirety of the relic neutralino surface, these decay leptons will be
soft.  For our analysis we include all production and decay
combinations based on the direct production $pp \to
\chi_i \chi_j$.  The backgrounds for the dilepton final state are
di-boson production dominated by $pp \to W^+W^-j$ production,
$\tau^+\tau^- j$ production, and $t\bar t$ production. For our
analysis we consider these three channels as the combined background.

Similar to the analysis in Section~\ref{sec:well-tempered} we generate
tree-level signal and background events with
\textsc{Madgraph5}~\cite{Alwall:2014hca},
\textsc{Pythia8}~\cite{Sjostrand:2007gs}, and
\textsc{Delphes3}~\cite{deFavereau:2013fsa} with the Snowmass detector
card~\cite{Anderson:2013kxz}. Jets are defined using the anti-$k_T$
jet algorithm~\cite{Cacciari:2008gp,Cacciari:2011ma} with $R=0.4$.  Since we will ask
for exactly one hard jet and veto events with additional hard jets, we
restrict the simulations to one hard jet in the matrix element.  As
generator-level cuts we ask for $p_{T,j} > 80~\gev$,
$p_{T,\ell}>5~\gev$, and for the backgrounds $\met > 480~\gev$.  For
the $t\bar{t}$ background we apply an anti-$b$-tag on the hardest jet,
conservatively implemented through a 20\% efficiency of passing the
anti-$b$-tag multiplied onto the $t\bar{t}$ cross section. Outside the
strongly boosted regime this detector acceptance for leptons limits us
to mass splittings above $5~\gev$. 

To reject the wide variety of different background processes we
require exactly one anti-$b$-tagged jet, sizeable missing transverse momentum,
and at least two isolated leptons ($e$ or $\mu$) with
\begin{alignat}{5}
p_{T,\ell} &= [10,50]~\gev &\qqquad |\eta_\ell| &< 2.5 &\qqquad m_{\ell \ell} &< m_{\ell \ell}^\text{max}  \notag \\
p_{T,j} &> 100~\gev &\qqquad |\eta_j| &< 2.5 &\qqquad \met &> 500~\gev. \; 
\label{eq:leptoncuts}
\end{alignat}
Events including additional jets with $p_{T,j} > 100~\gev$, mostly
coming from $t\bar{t}$ production, are vetoed. 
The two highest-$p_T$ leptons are selected from events
with more than two leptons. The reasoning behind
the stiff cut on missing transverse momentum again follows
Figure~\ref{fig:boostelectro}.  We could increase it to $\met >
1~\tev$, which increases the signal-to-background ratio $S/B$ but
reduces $S/\sqrt{B}$. In terms of the reach of a $100~\tev$ collider
the two scenarios are roughly equal. Finally, we could ask for a
harder jet without a large effect on the signal rate, but unlike in
Section~\ref{sec:well-tempered} the two leptons in this signature,
combined with very large missing momentum, should guarantee an
efficient trigger.

For all processes listed in Eq.\eqref{eq:process2} we know that the
$\chi^0_2$ and $\chi^\pm_1$ decays involve off-shell bosons with soft
leptons. In contrast the background leptons tend to come from on-shell
gauge boson decays. In addition to the individual upper limit on
$p_{T,\ell}$ we can further reduce the $WW/ZZ$ and $t\bar{t}$
backgrounds using an upper limit on $m_{\ell \ell}$ for the two
hardest leptons~\cite{Han:2014kaa}. Because of the number of distinct signal
paths \eqref{eq:process2} and the variety of mass splittings over the relic surface, 
the optimal cut for $m_{\ell \ell}$ varies, often over values such that $m_{\ell \ell} < 90 ~\gev$. 
This is why after all other cuts, we apply the optimal value for an $m_{\ell \ell}$ cut determined
individually for each parameter point and chosen to maximize the
statistical significance $S/\sqrt{B}$. In cases where the $m_{\ell
  \ell}$ cut does not improve significance, for example due to too
small signal rates, this procedure will automatically remove the cut
altogether.

%---------------------------------------
\begin{figure}[t]
\begin{center}
  \includegraphics[width=0.60\textwidth]{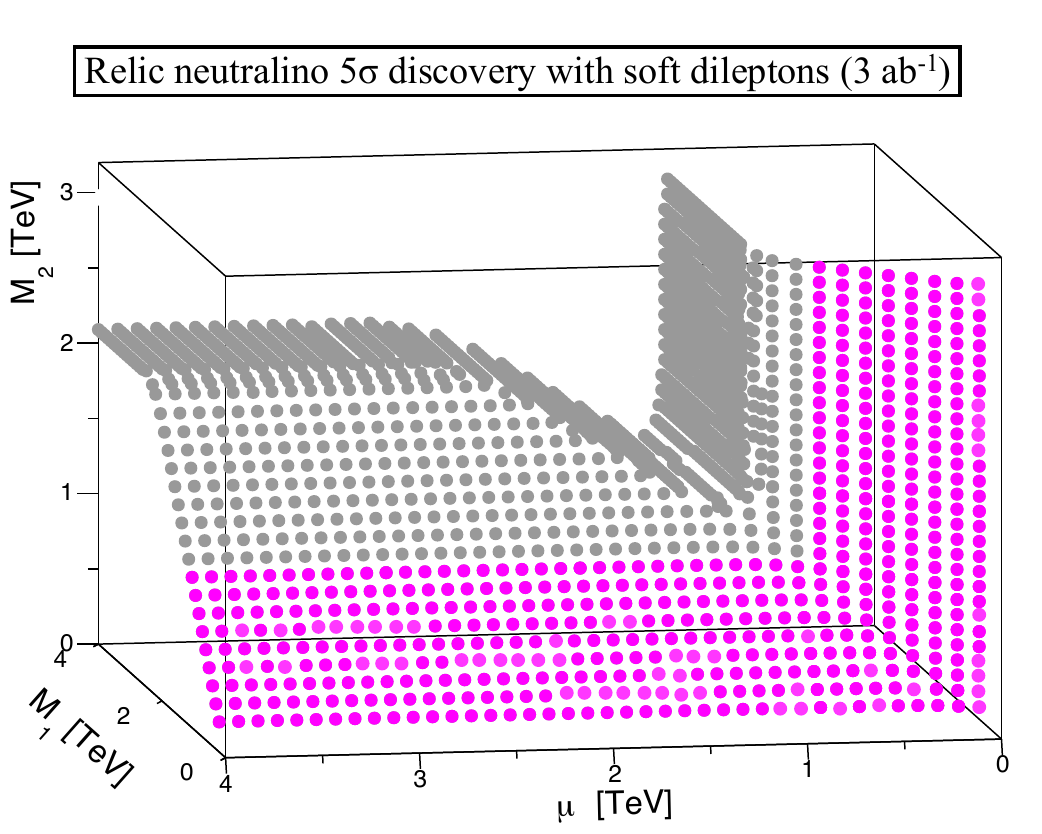}
\end{center}
  \caption{Discovery range of the dilepton signal,
    defined by $S/\sqrt{B} > 5$ at $3~\iab$.  The colored points can be discovered with no more than $3~\iab$ of data, while the grey points will require more luminosity.}
  \label{fig:lepton_reach}
\end{figure}
%---------------------------------------

The two limiting factors for this analysis are the signal production
cross section shown in Figure~\ref{fig:surface_csx} and the mass
splitting shown in Figure~\ref{fig:surface_split}.  In
Figure~\ref{fig:lepton_reach} we show the $5\sigma$ discovery
range with an integrated luminosity of $3~\iab$. For the points in
reach of a $100~\tev$ collider we find typical signal--to--background
ratios $S / B \gtrsim 1/5$. As mentioned above, the balance between
statistical and systematic uncertainties can be balanced by the choice
of $\met$ cut. The significance and the signal-to-background ratio
worsen on the bino--wino surface towards higher $M_1$ and on the
bino--higgsino surface towards higher $\mu$, in both cases
corresponding to a decreased mass splitting.

%%%%%%%%%%%%%%%%%%%%%%%%%%%%
\section{Conclusions}
\label{sec:conclusions}
%%%%%%%%%%%%%%%%%%%%%%%%%%%%

In this article we have mapped the surface of bino-wino-higgsino parameter
space which provides the correct relic abundance, assuming all other
superpartners are decoupled. The majority of this relic neutralino
surface is compressed: the mass splitting between the NLSP and LSP is
typically much less than the mass of electroweak bosons. Furthermore, large portions
of the relic MSSM neutralino surface, in particular, the
bino-wino portion, will be inaccessible to all planned direct
detection searches. This makes
the surface a ripe target at future colliders for dark matter searches
that require large missing transverse momentum and associated
electroweak particles with low momenta.

%---------------------------------------
\begin{figure}[t]
\begin{center}
  \includegraphics[width=0.60\textwidth]{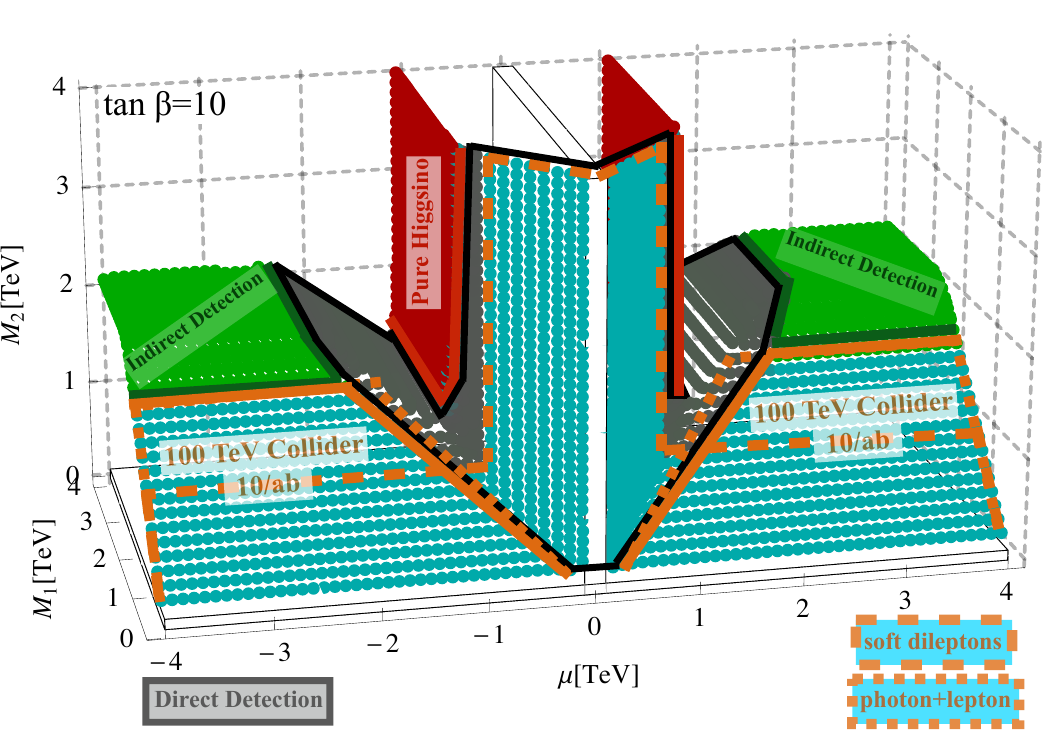}
\end{center}
  \caption{Relic neutralino surface divided into prospective regions of future direct, future indirect, and collider detection accessibility as indicated. Direct detection regions were determined by requiring the spin-independent cross-section be $> 10^{-45} ~\rm{cm}^2$, as calculated at tree level by \textsc{micrOmegas3}. Regions detectable by a 100 TeV collider using missing transverse momentum, a hard jet, and either soft dileptons or a lepton and photon are marked with long and short-dashed lines, respectively. The white box occludes regions already ruled out by LEP.}
  \label{fig:prospects}
\end{figure}
%---------------------------------------

To discover relic neutralinos at a next-generation $100~\tev$
collider, we have demonstrated two effective methods: searches for large missing transverse momentum, a jet from initial state
radiation, and either a soft photon and lepton or two soft leptons.
These final states are produced in the decays of slightly heavier electroweakinos down to the LSP. Our searches especially benefit from the increased center of
mass energy at a $100~\tev$ collider as we can look for boosted electroweakinos. Once boosted, the electroweakino signal induces large missing energy, but relatively soft leptons, as their energy is set by the inter-electroweakino
mass splittings. This combination of large missing energy and soft leptons is difficult to achieve in the SM where both the $\slashed p_T$ and leptons come from the same object, typically an electroweak gauge boson.

Quantitatively, we have shown that the photon and lepton search will
discover up to $1.5~\tev$ relic bino-wino dark matter with less than
ten inverse attobarns of luminosity. Assuming a next-generation
detector sensitive to photons with momenta as small as $5~\gev$, we
find that neutralino mass splittings as small as $1~\gev$ can be
discovered in the boosted environment present at a $100~\tev$ collider. We also showed that a
$100~\tev$ collider is particularly adept at finding relic bino-winos
and bino-higgsinos using the soft dileptons emitted in heavier neutralino and
charging decays, for relic masses up to $\sim \tev$. While this leaves
some bino-higgsino and wino-higgsino portions of the surface
un-discoverable at a $100~\tev$ collider, this same relic neutralino
parameter space is the most accessible to next generation direct
detection experiments, as illustrated in Figure \ref{fig:prospects}. We conclude that a $100~\tev$ collider will
be a spectacular complement to other dark matter searches and
often provides the best prospects for dark matter discovery, both over the MSSM
relic neutralino surface and for electroweak dark matter states with
$1-50~\gev$ mass splittings. 

\acknowledgments

We would like to thank Vera Gluscevic, Rafael Lang, Annika Peter, and
Pedro Schwaller for discussions. JB is grateful to the CERN theory
group for their hospitality while this paper was finalized.  TP would
like to thank the Fermilab Theory Group, where this project once
started. TS acknowledges support by the IMPRS for Precision Tests of
Fundamental Symmetries. The work of AM was partially supported by the
National Science Foundation under Grant No. PHY-1417118. This research
was supported in part by the Notre Dame Center for Research Computing
with computing resources. This work was partially supported by World Premier International Research Center Initiative (WPI), MEXT, Japan. Fermilab is operated by Fermi Research Alliance, LLC under Contract No. DE-AC02-07CH11359 with the United States Department of Energy.

\bibliographystyle{JHEP}
\clearpage
\bibliography{nimatronprd}

\end{document}